\documentclass[journal,comsoc]{IEEEtran}
\usepackage[T1]{fontenc}

\ifCLASSINFOpdf
  \usepackage[pdftex]{graphicx}
  \graphicspath{{figures/}}
  \DeclareGraphicsExtensions{.pdf,.jpeg,.png}
\else
\fi
\usepackage{amsmath}
\usepackage[cmintegrals]{newtxmath}
\ifCLASSOPTIONcompsoc
  \usepackage[caption=false,font=normalsize,labelfont=sf,textfont=sf]{subfig}
\else
  \usepackage[caption=false,font=footnotesize]{subfig}
\fi
\newtheorem{thm}{Theorem}
\newtheorem{defn}{Definition}
\newtheorem{exmp}{Example}
\newtheorem{cor}{Corollary}
\newtheorem{lem}{Lemma}

\newcommand\tab[1][0.6cm]{\hspace*{#1}}

\begin{document}
%
\title{A Survey on Trapping Sets and Stopping Sets}
%
%
%

\author{Aiden~Price
        and~Joanne~Hall, \textit{Member, IEEE}.
\thanks{A. Price is with the Science and Engineering Faculty, Queensland University of Technology, Queensland,
QLD, Australia. e-mail: a11.price@qut.edu.au.}
\thanks{J. Hall is with the School of Science, Royal Melbourne Institute of Technology, Melbourne.}
\thanks{Manuscript received MO DATE, YEAR; revised MO DATE, YEAR.}}

%
%

\markboth{ May 2017}%
{Shell \MakeLowercase{\textit{et al.}}: Bare Demo of IEEEtran.cls for IEEE Communications Society Journals}
%



\maketitle

\begin{abstract}
LDPC codes are used in many applications, however, their error correcting capabilities are limited by the presence of stopping sets and trappins sets. Trappins sets and stopping sets occur when specific low-wiehgt error patterns cause a decoder to fail. Trapping sets were first discovered with investigation of the error floor of the Margulis code. Possible solutions are constructions which avoid creating trapping sets, such as progressive edge growth (PEG), or methods which remove trapping sets from existing constructions, such as graph covers. This survey examines trapping sets and stopping sets in LDPC codes over channels such as BSC, BEC and AWGNC.
\end{abstract}


%
\IEEEpeerreviewmaketitle

\section{Introduction}
%
%
%
%
\IEEEPARstart{A}{s} technology advances, we wish to communicate over longer distances and have the ability to stay connected even over poor communication channels. While quasi-cyclic low-density parity-check (QC-LDPC) codes are one of the best ways to achieve this \cite{2DDOV2016}, their performance in many cases is limited by the presence of trapping sets and stopping sets. Trapping sets and stopping sets can cause iterative decoding methods to fail with relatively few errors. Finding ways to avoid or remove trapping sets and stopping sets will further improve the already high performance of LDPC codes and bring their performance curves even closer to the Shannon limit \cite{MN1996}, \cite{S1948}.

Performance optmization is becoming incresingly crucial as the world moves further into the digital age. An increase in the speed at which digital communication occurs through modern applications such as WiFi \cite{802.11n2009WiFi} and DVB-S2 \cite{302307v1.3.1ETSI2013} has drastic implications on the overall productivity of the world.

In 1962, Gallager \cite{G1962} introduced low-density parity-check codes (LDPCs). LDPC codes are a class of binary linear block codes with a sparse parity-check matrix. An advantage of using LDPC codes is that they are able to provide error control which is very close to the capacity for many different channels \cite{MN1996}. This categorizes LDPC codes as one of few capacity-approaching codes; error correction methods which can allow the noise in a channel to be set very close to its theoretical maximum while maintaining error-correcting ability \cite{BCH2011}.

The performance of an error correction method is based upon two properties; the performance of such a code over a channel with variable noise and the optimal bit error ratio (BER) of a code with sufficient signal-to-noise ratio (SNR). The optimal BER is known as the error floor of a code \cite{R2003}, and is discussed in different papers in terms of bit error rate (BER), frame error rate (FER), block error rate and symbol error rate, depending on the application being addressed (see \cite{JW2004}, \cite{ICV2008} for examples of such error floor analysis). Consideration of the error floor is one of the most important aspects of constructing a high-performing LDPC code \cite{ICV2008}.

Analysis of the performance of LDPC codes over the binary erasure channel (BEC) led to the discovery of stopping sets in 2002 \cite{DPTRU2002}. The Margulis construction \cite{Margulis1982} improved upon the performance of Gallager codes, though a weakness in this construction led to a high error floor over the additive white Gaussian noise channel (AWGNC) compared to the performance of other constructions of the time \cite{MP2003}. This high error floor was due to the presence of stopping sets. Stopping sets over BEC as described in \cite{DPTRU2002} became a well understood problem and led to the definition of trapping sets, which are defined over AWGNC and BSC. In some early works trapping sets are called near-code words \cite{R2003}, \cite{MP2003}.

Trapping sets and stopping sets are an important topic, worthy of a stand-alone survey.

\section{Preliminaries and Notation}
In order to engage with the literature on stopping sets and trapping sets an overview of the preliminaries is necessary. We provide a short review of the literature surrounding LDPC codes, the common transmission channels, and common decoding techniques.

\begin{defn}\label{def: LC}{\normalfont}\cite{B2014}
A binary $[n, k, d]$ \textbf{linear code}, $\mathcal{C}$, is a $k$-dimensional subspace of an $n$-dimensional vector space, $\mathbb{F}_{2}^{n}$ which is used to provide structure to a message vector for transmission over a channel.
\end{defn}

In order to transmit messages over communication channels using an error correcting code, we encode the message using a generator matrix.

\begin{defn}\label{def: GM}{\normalfont}\cite{BCH2011}
The \textbf{generator matrix}, \textbf{G}, of a code, $\mathcal{C}$, is a matrix which has dimensions $k \times n$. The $k$ rows of $\mathbf{G}$ correspond to linearly independent code words which form a basis of $\mathcal{C}$.
\end{defn}

One of the most important aspects of error correction is the process of decoding. A parity-check matrix allows us to identify whether errors have been introduced during transmission. This matrix can also be represented by a Tanner graph.

\begin{defn}\label{def: PCM}{\normalfont}\cite{BCH2011}
A \textbf{parity-check matrix}, $\mathbf{H}$, of $\mathcal{C}$ is a matrix which generates the nullspace of the code. This means that a code word, $\mathbf{c}$, is in the code, $\mathcal{C}$, iff $\mathbf{H}\cdot\mathbf{c}^{T} = \mathbf{0}$, where $\mathbf{0}$ is an $r \times 1$ null vector. $\mathbf{H}$ has dimensions $(n-k) \times n$.
\end{defn}

\begin{figure*}[!t]
\centering
\subfloat[][]{\includegraphics[width=1.44in]{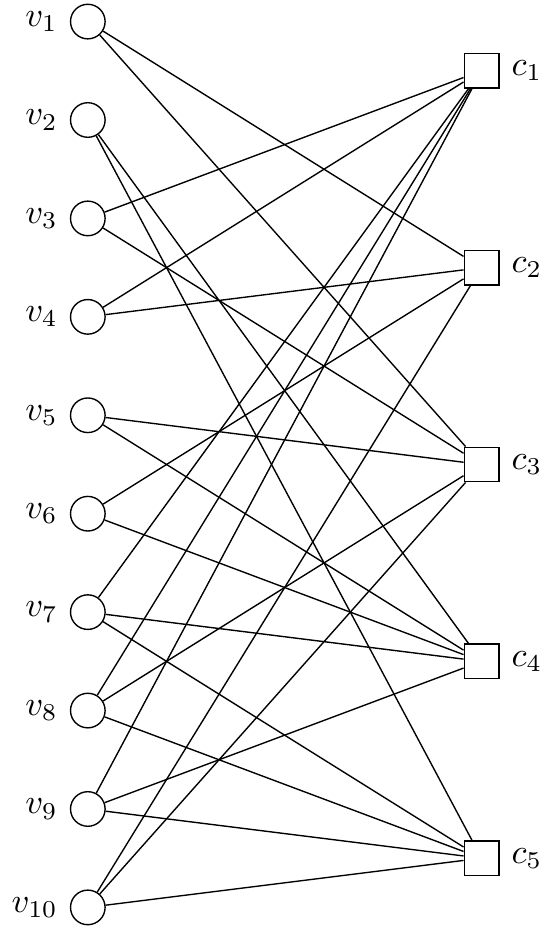}}
\hspace{1pt}\vrule\hspace{1pt}
\subfloat[][]{\includegraphics[width=1.38in]{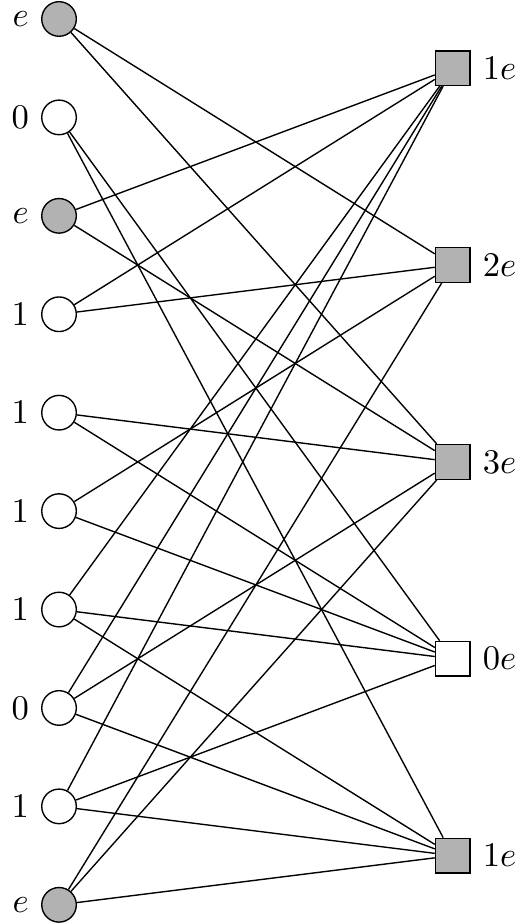}}
\raisebox{18\height}{$\rightarrow$}
\subfloat[][]{\includegraphics[width=1.38in]{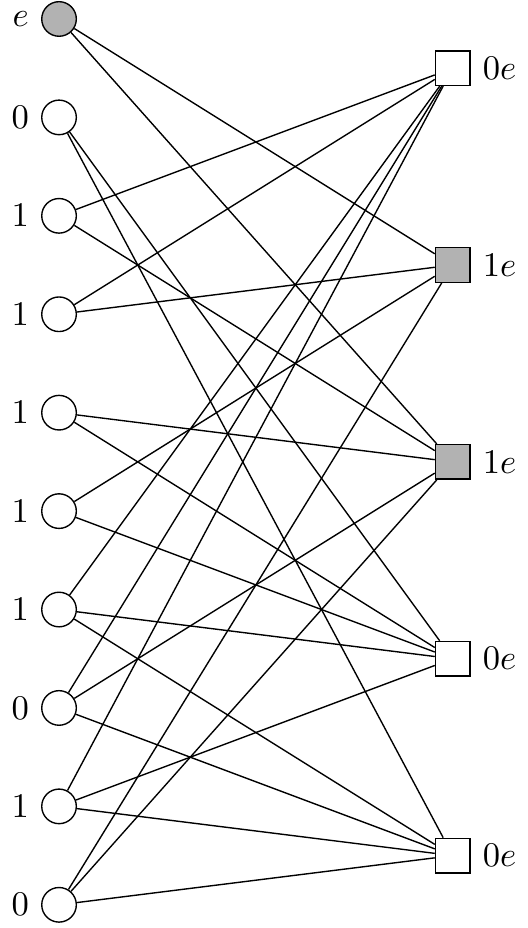}}
\raisebox{18\height}{$\rightarrow$}
\subfloat[][]{\includegraphics[width=1.38in]{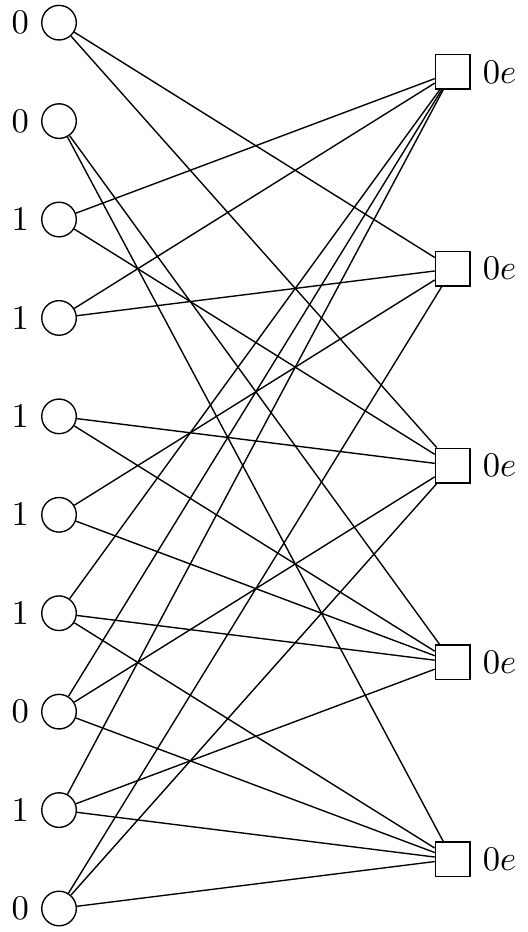}}
\caption{A $5 \times 10$ irregular Tanner graph (a) used to demonstrate the ER decoding algorithm with the received vector $v = [e,0,e,1,1,1,1,0,1,e]$ over BEC. Nodes of interest are highlighted gray. (b) shows steps 1 and 2 of the ER algorithm's first iteration. (c) shows the changes made in step 3 and then step 2 of the second iteration. (d) shows the changes made in step 3 again, this time revealing that no further erasures exist. The algorithm then terminates on step 4 of this iteration, thus successfully correcting the received vector in 2 iterations.}
\label{ER Tanner}
\end{figure*}

\begin{defn}\label{def: TG}{\normalfont}\cite{B2014}
The parity-check matrix, $\mathbf{H}$, may be represented by a bipartite graph with variable node set, $V$, and check node set, $C$. This bipartite graph is denoted $G(\mathbf{H}) = (V \cup C, E)$, where the columns of $\mathbf{H}$ indicate the variable nodes in $V$ and the rows of $\mathbf{H}$ indicate the check nodes in $C$. For $i \in V$ and $j \in C$, $(i, j) \in E$ if and only if $H_{ij} = 1$. This bipartite graph is known as a \textbf{Tanner graph} with $r = n-k$ check nodes and $n$ variable nodes (see Fig. \ref{ER Tanner}).
\end{defn}

We can also refer to individual variable nodes; let the $i$-th variable node be $v_{i}$ and the $j$-th check node be $c_{j}$ \cite{Richter2006}.

\begin{defn}\label{def: LDPC}{\normalfont}\cite{MM2010}
If a parity-check matrix of a code is sparse, then the corresponding code, $\mathcal{C}$, is called a \textbf{low-density parity-check (LDPC) code}.
\end{defn}

We note that the classification of \textbf{sparse} used in the context of LDPC codes is that there are fewer ones in the parity-check matrix than there are zeros \cite{BCH2011}. The sparse nature of LDPC codes means that decoding processes have a fast run-time, as there are fewer operations to compute when compared to a non-sparse parity-check matrix. Two more important features of Tanner graphs are neighbours and node degrees.

\begin{defn}\label{def: NND}{\normalfont}\cite{MM2010}\cite{SS2012}
For a variable node $v_{i}$ and check node $c_{j}$, if $(i,j) \in E$ we say that nodes $v_{i}$ and $c_{j}$ are \textbf{neighbours}. The \textbf{degree} of a node in the Tanner graph is defined as the number of edges it is connected to.
\end{defn}

From the node degree definition we can also define regular LDPC codes.

\begin{defn}\label{def: RLDPC}{\normalfont}\cite{G1962}
An LDPC code is called $(d_{v},d_{c})\mathbf{-regular}$ if each variable node, $v$, has degree $d_{v}$ and each check node, $c$, has degree $d_{c}$. We denote an LDPC code of this form a $\mathcal{C}(d_{v},d_{c})$ code.
\end{defn}

LDPC codes are designed to be used as error-correction methods over a variety of communication channels. There are three communication channels discussed in this paper; the binary erasure channel (BEC), the binary symmetric channel (BSC) and the additive white gaussian noise channel (AWGNC) \cite{H1986}. Though these channels handle data transmission in different ways, their encoding and decoding goals are the same.

\subsection{Encoding and Verification}
The process of transforming a message vector into its associated code word is known as encoding. Every code word $\mathbf{c} = [c_{0},c_{1},...,c_{n-1}] \in \mathcal{C}$ can be expressed as
\begin{equation*}
\mathbf{c} = \mathbf{m}\cdot\mathbf{G}.
\end{equation*}
Where $\mathbf{m} = [m_{0},m_{1},...,m_{k-1}]$ is the mesage vector \cite{BCH2011}. The code word, $\mathbf{c}$, has the original $k$ information bits as well as an additional $r$ parity bits to give the code word a length of $n$ bits. As the parity-check matrix, $\mathbf{H}$, is the nullspace of the code, we can use it as a verification method to test if a recieved vector is a code word \cite{B2014}. The product $\mathbf{H}\cdot\mathbf{c}^{T}$ is denoted the \textbf{syndrome}, $\mathbf{s}$, of $\mathbf{c}$ through $\mathbf{H}$. For a given vector, $\mathbf{v}$,
\begin{equation*}
\mathbf{v} \in \mathcal{C} \text{ iff } \mathbf{s} = \mathbf{H}\cdot\mathbf{v}^{T} = \mathbf{0}.
\end{equation*}
There must be an even number of ones in the components of the product $\mathbf{H}\cdot\mathbf{c}^{T}$ which add to give $\mathbf{s} = 0$. This is known as the even-parity constraint \cite{MN1996}. After a code word, $\mathbf{c}$, is transmitted through a channel, the other party receives a vector, $\mathbf{v}$. If $ \mathbf{s} = \mathbf{H}\cdot\mathbf{v}^{T} \neq \mathbf{0}$, then $\mathbf{v} \notin \mathcal{C}$ and so we use error correcting techniques in attempt to correct $\mathbf{v}$ and recover $\mathbf{c}$.

The upper bound on the error correcting ability of an LDPC code is determined by the minimum distance of the code. In order to define the minimum distance of a code, we will first define Hamming weight and Hamming distance.

\begin{defn} \cite{PW1972}
The \textbf{Hamming weight}, $w$, of a vector is the number of its non-zero elements. The Hamming weight of a binary vector is therefore the number of ones in the vector. The \textbf{Hamming distance}, $d$, between two vectors, $\mathbf{x}$ and $\mathbf{y}$, is the number of places in which they differ; written as $d(\mathbf{x},\mathbf{y})$.
\end{defn}

In the literature, Hamming weight and Hamming distance are often referred to using the terms ``weight'' and ``distance'' \cite{B2014}. The weight of code words affects the number of operations performed in decoding and the distance between code words affects how many errors can be corrected.

\begin{defn} \cite{H1986}
The \textbf{minimum distance of a code}, $\mathcal{C}$, is defined as the smallest Hamming distance between any two code words in the code,
\begin{equation*}
d(\mathcal{C}) = min \{d(\mathbf{x},\mathbf{y}) \, | \, \mathbf{x},\mathbf{y} \in \mathcal{C}, \mathbf{x} \neq \mathbf{y}\}.
\end{equation*}
\end{defn}

The following theorem and corollary describe a code's error detection and correction abilities using minimum distance \cite{H1986}.

\begin{thm}{\normalfont\cite{H1986}} A code $\mathcal{C}$ can detect up to $s$ errors in any code word if $d(\mathcal{C}) \geq s + 1$. A code $\mathcal{C}$ can correct up to $t$ errors in any code word if $d(\mathcal{C}) \geq 2t + 1$.
\end{thm}

\begin{cor}{\normalfont\cite{H1986}} If a code $\mathcal{C}$ has minimum distance $d$, then $\mathcal{C}$ can be used to either detect up to $d-1$ errors, or to correct up to $(d-1)/2$ errors in any code word.
\end{cor}

If the minimum distance of a code is too small, then it cannot provide sufficient error correction. This is demonstrated in Example \ref{ex:distance}

\begin{exmp} \label{ex:distance}
Let two code words be given as;
\begin{align*}
c_{1} &= [0 1 0 0 0 0 1 1 1 1] \\
c_{2} &= [0 1 0 0 1 1 1 0 1 0].
\end{align*}
The distance between these codewords is given as $d(c_{1},c_{2}) = 4$. Take the following error vector:
\begin{align*}
e &= [0 0 0 0 1 1 0 1 0 1].
\end{align*}
If $e$ is added to $c_{1}$ then the resulting code word is identical to $c_{2}$, demonstrating the importance of minimum distance.
\end{exmp}

The minimum distance of an LDPC code is also related to its code-rate; a large code rate lowers the upper bound on the minimum distance of a code.

\begin{defn}
The \textbf{code rate}, $R(\mathcal{C})$, of an LDPC code is the portion of information bits sent in comparison to the entire code vector sent, written as
\begin{equation*}
R(\mathcal{C}) =\frac{k}{n},
\end{equation*}
where $0 \leq R(\mathcal{C}) \leq 1$.
\end{defn}

The code rate and minimum distance often determine the error correcting capability of an LDPC code, though the decoding algorithm plays a direct part in the time it takes to decode messages.


\subsection{Communication Channels and Decoding Basics}
The communication channel by which transmission occurs impacts the error correction algorithms that are chosen. A communication channel can be modelled as a triple which contains an input alphabet, an output alphabet and the probability of transition between a symbol in the input alphabet and a symbol in the output alphabet \cite{S1948}.

The binary erasure channel (BEC) is one of the simplest, non-trivial channel models \cite{RU2008},\cite{Elias1955}.

\begin{defn}\cite{Poddar2007},\cite{MM2010}
The \textbf{Binary Erasure Channel (BEC)} is a communication channel with two input symbols, $0$ and $1$, and three output symbols, $0$, $1$ and $e$ (the erasure symbol). The BEC has an erasure probability, $p$, where given an input, $c_{i}$, the output, $v_{i}$, is defined by the probability formulae $P[v_{i} = c_{i}] = 1-p$ and $P[v_{i} = e] = p$.
\end{defn}

Analysis of the BEC significantly advanced modern understanding of error correction \cite{RU2008}. An example of a simple decoding process over the BEC is the Edge Removal algorithm.

\begin{defn} \cite{MM2010} \label{ERalg}
Let $\mathbf{c}\in\mathcal{C}\subseteq\{0,1\}^{n}$ be a binary code word transmitted over the BEC and $\mathbf{v}\in\{0,1,e\}^{n}$ be the received vector. The \textbf{Edge Removal (ER)} algorithm proceeds as follows:

\begin{enumerate}
\item Initial Step: The value of each received vector bit, $v_{i}$ is assigned to each variable node $i \in V$ of the Tanner Graph.
\item The check nodes $c_{i}\in C$ count the number of erased bits which are neighbours in the Tanner graph, $G(H)$.
\item If check node $c_{i}$ neighbours only one $e$ symbol in $v$, the even parity constraint uniquely determines the original value of $e$ for that variable node.
\item Repeat steps (2) and (3) until either all erasures have been recovered or until every check node that is a neighbour of an erased bit is a neighbour of at least two erased bits.
\end{enumerate}
\end{defn}

For step (4) above, if the latter occurs, then the decoder has failed due to the presence of a stopping set (see Section \ref{SSoverBEC}).

We provide an example of the ER decoding process in Fig. \ref{ER Tanner}, where $\circ$ represents a variable node and $\square$ a check node. This example is found in \cite{Richter2006}.

For this decoding example, we use an irregular $5 \times 10$ parity-check matrix, $\mathbf{H}$ and demonstrate the decoding of received vector $v = [e,0,e,1,1,1,1,0,1,e]$ using the ER algorithm over the BEC.

\begin{figure*}[!t]
\centering
\subfloat[][]{\includegraphics[width=1.45in]{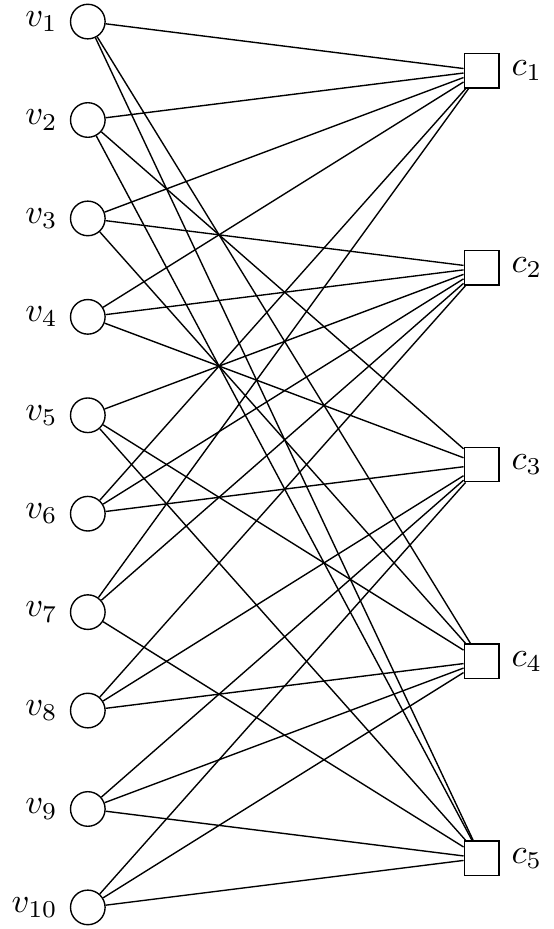}}
\hspace{1pt}\vrule\hspace{1pt}
\subfloat[][]{\includegraphics[width=1.34in]{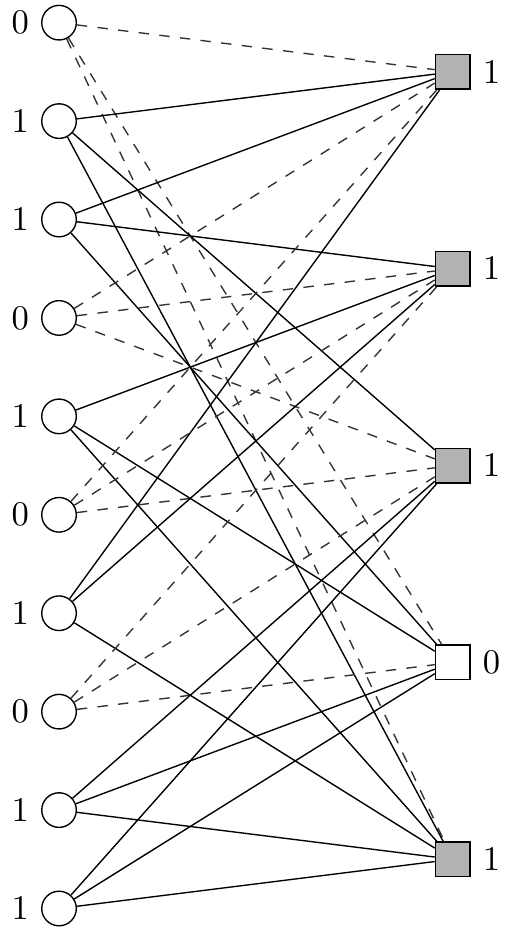}}
\raisebox{18\height}{$\rightarrow$}
\subfloat[][]{\includegraphics[width=1.51in]{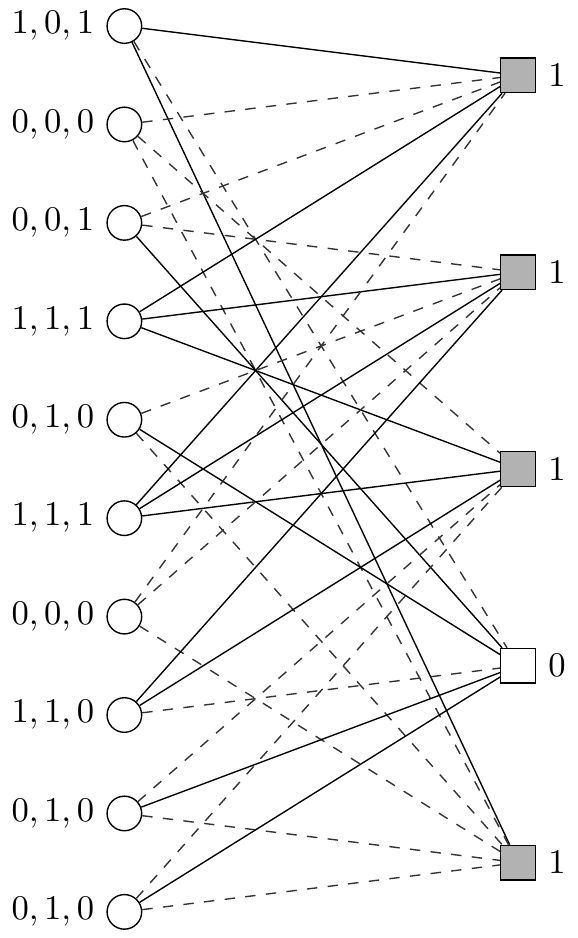}}
\raisebox{18\height}{$\rightarrow$}
\subfloat[][]{\includegraphics[width=1.51in]{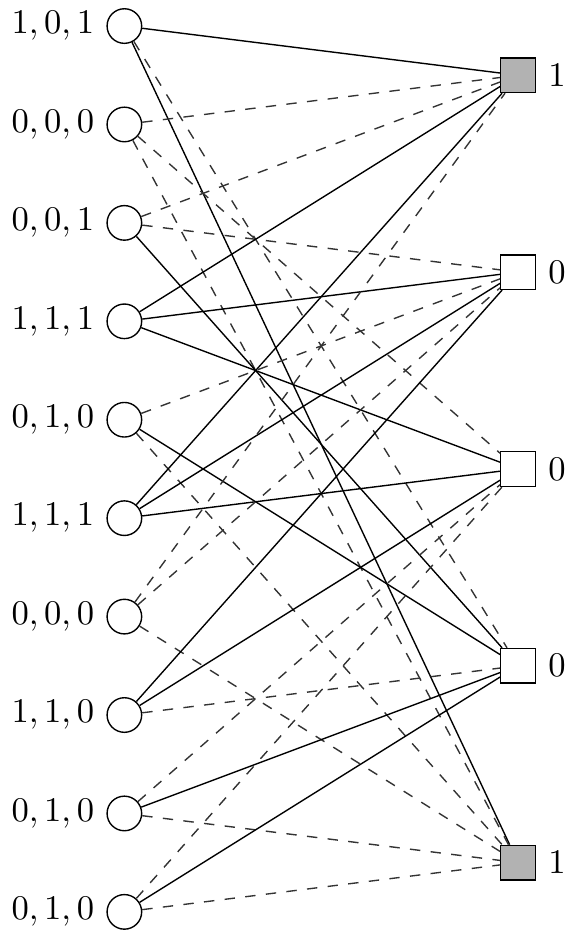}}
\caption{A $5 \times 10$ (6,3)-regular Tanner graph (a) used to demonstrate the Gallager A decoding algorithm with the received vector $v = [0,1,1,0,1,0,1,0,1,1]$. We represent a 0 being sent along an edge by a dashed line and a 1 with a full black edge. (b) shows step 1 of the Gallager A algorithm, as well the check node calculation, taken as the addition mod 2 of all incoming message from variable nodes adjacent to each check node (denoted $v_{i} \rightarrow c_{i}$). (c) then shows step 2, $c_{i} \rightarrow v_{i}$. Lastly, (d) shows step 3, $v_{i} \rightarrow c_{i}$. Due to the complexity of this algorithm, only one full iteration has been shown, though step 4 in Definition \ref{GAalg} describes how decoding continues.}
\label{GA Tanner}
\end{figure*}

Another communication channel is the binary symmetric channel (BSC).

\begin{defn}\cite{MM2010}, \cite{Poddar2007}
The \textbf{Binary Symmetric Channel (BSC)} is a communication channel with two input symbols, $0$ and $1$, and two output symbols, also $0$ and $1$. The BSC has an error probability, $p$, where given an input, $c_{i}$, the output, $v_{i}$, is defined by the probability formulae $P[v_{i} = c_{i}] = 1-p$ and $P[v_{i} = \bar{\mathbf{c}_{i}}] = p$.
\end{defn}

An example of a decoding process over the BSC is the Gallager A algorithm \cite{G1962}.

\begin{defn} \cite{G1962}, \cite{Poddar2007}, \cite{Shokrollahi2003}
\label{GAalg}
Let $\mathbf{c}\in\mathcal{C}\in\{0,1\}^{n}$ be a binary code word transmitted over the BSC and $\mathbf{v}\in\{0,1\}^{n}$ be the received vector. The \textbf{Gallager A} algorithm proceeds as follows:
\begin{enumerate}
\item Initial Step: The value of each received vector bit, $v_{i}$ is assigned to each variable node $i \in V$ of the Tanner graph.
\item After this, a check node $c_{i}$ sends to all neighbouring variable nodes $v_{i}, \dots, v_{j}$ the sum (mod 2) of all of the adjacent variable nodes except for the node itself (where $j$ is the degree of check node $c_{i}$).
\item Each variable node, $v_{i}$ then sends the following to their adjacent check nodes: If all messages from check nodes $c_{i}$ other than the target check node of the message are equal, then $v_{i}$ sends that message back, otherwise it resends its prior value.
\item Repeat steps (2) and (3) until either all variables nodes send the same values over two consecutive iterations or when a pre-set max iteration count is reached.
\end{enumerate}
\end{defn}

The \textbf{Gallager B} algorithm offers improved decoding with an additional step on each loop within the algorithm \cite{Shokrollahi2003}. For each degree, $j$, and each check node loop, $i$, there is a pre-chosen threshold value, $b_{i,j}$. Throughout the steps involved in check node $c_{i}$ for each variable node, $v$, and each adjacent check node, $c$, if at least $b_{i,j}$ neighbours of $v$ excluding $c$ sent the same information in the previous round, then $v$ sends that information to $c$; otherwise $v$ sends its received value to $c$. Algorithm A is a special case of algorithm B, where $b_{i,j} = j - 1$ independent of the round \cite{Shokrollahi2003}.

Throughout the decoding procedure, if the pre-set max iteration count is reached without completion, the decoder has failed due to the existence of a trapping set (see Section \ref{TSoverOther}). An example of the first steps of the Gallager A algorithm (see Fig. \ref{GA Tanner}) demonstrates the differences between the decoding considerations made between the BEC and the BSC.

The most complex channel considered here is the binary input additive white gaussian noise channel, expressed commonly either as the BI-AWGNC or just as the AWGNC \cite{TKJ2007}.

\begin{defn} \cite{TKJ2007}
\label{defn:AWGN}
Let $X \in \{0,1\}^{*}$ be a message vector, where $^{*}$ denotes an arbitrary length. The \textbf{additive white Gaussian noise channel (AWGNC)} maps the input vector, $X$, to the vector $X' \in \{+1,-1\}^{*}$ and then adds the result with Gaussian white noise to give an output vector $Y = X' + W$, where $W \sim \mathcal{N}(0,N_{0}/2E_{b})$.
\end{defn}

Each code symbol, $y \in Y$, carries with it a signal to noise ratio (SNR) of $E_{b}/N_{0}$ and the conditional distribution of $Y$ is
\begin{equation}
P(y | x') = P_{W}(y-x') = \frac{1}{\sqrt{2\pi(N_{0}/2E_{b})}}\cdot exp\bigg(\frac{-(y-x')^{2}}{(N_{0}/E_{b})}\bigg),
\end{equation}
which gives the output alphabet for the AWGNC as $y \in \mathbb{R}$. As in the BEC and BSC, we would like to have some indication of the errors that a channel is introducing to the code word. A metric used for the AWGNC is the log likelihood ratio (LLR).
\begin{equation}
L(y|x) = ln\bigg(\frac{P(y|x = 0)}{P(y|x = 1)}\bigg).
\end{equation}

This $L$-value describes the likelihood that $x$ is $0$ or $1$. If $L$ is positive then $P(y|x = 0) > P(y|x = 1)$ and thus the input estimate should be $\hat{x} = 0$. There are methods to map from $Y \in \mathbb{R}$ to $Y' \in \{0,1\}^{*}$ and, as such, all decoding methods used over the BSC can be implemented on the AWGNC. However, high performing decoding algorithms, such as maximum-likelihood decoders, the sum-product algorithm and the max-product algorithm \cite{R2003},\cite{CG2005},\cite{SLG2004} utilize the LLR information to improve decoding speed \cite{HXMAH2015}. The BSC can be used for these channels as LLR values are defined over the BSC, though the AWGNC more closely models the influence of real-world communication channels and is favoured for high performance simulations \cite{BCH2011},\cite{TKJ2007}.

\begin{exmp}
The BSC has a conditional LLR function as the bit flipping probabilities are well understood for the outputs
\begin{equation}
L_{BSC}(y|x) = 
\begin{cases} 
      ln(\frac{1-p}{p}) & y = 0 \\
      ln(\frac{p}{1-p}) & y = 1
   \end{cases}
\end{equation}
As the noise determines the values of $y$ in the AWGNC and $y \in \mathbb{R}$, the LLR on this channel is defined as
\begin{equation}
L_{AWGNC}(y|x) = 4\frac{E_{b}}{N_{0}}y.
\end{equation}
\end{exmp}

The decoding algorithms used on the AWGNC are far more complex than those on the BEC and BSC and, as such, we provide an overview of various methods rather than detailed definitions and examples. Decoding methods used over the AWGNC tend to be message passing algorithms, where nodes send information to their neighbours to correct errors based on the structure of the parity-check matrix \cite{SLG2006}. The original message-passing algorithm \cite{G1962} is an example of a flooding schedule \cite{KFL2001} where in each iteration, all variable nodes and subsequently all check nodes pass new messages to their neighbours. Another example of a flooding schedule is the sum-product algorithm (see \cite{CG2005}).

\begin{figure}[!t]
\centering
\subfloat[][]{\includegraphics[width=1.34in]{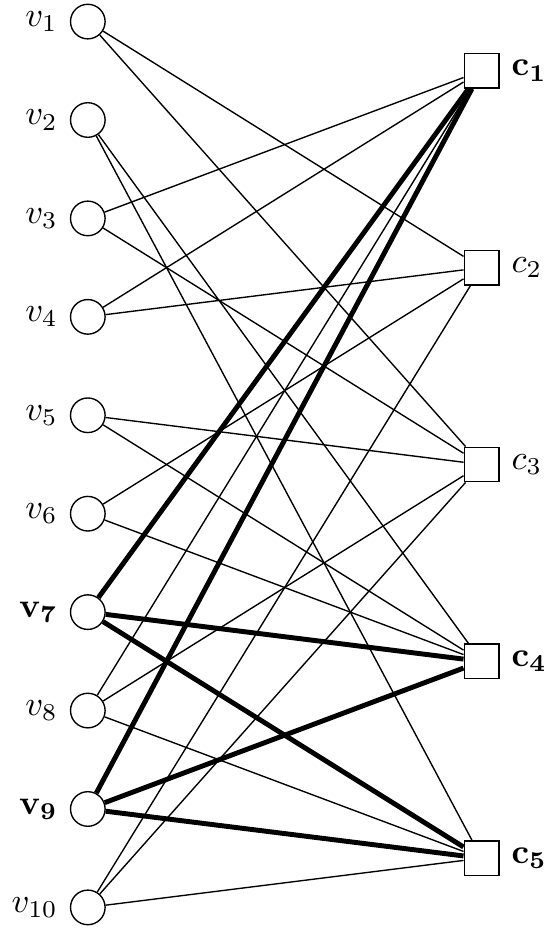}}
\raisebox{18\height}{$\rightarrow$}
\subfloat[][]{\raisebox{0.65\height}{\includegraphics[width=1.34in]{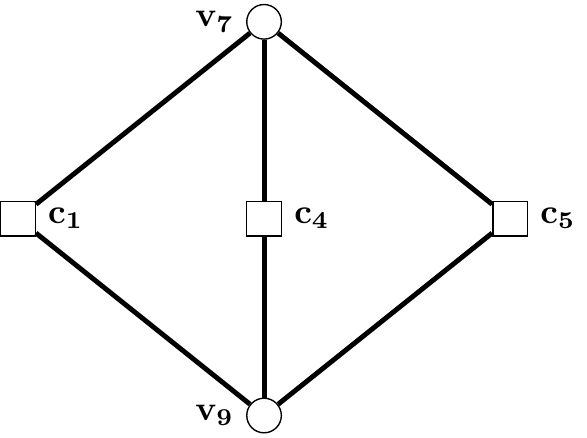}}}
\caption{(a) Tanner graph for the irregular $5 \times 10$ parity-check matrix given in Example \ref{SSvisual}. (b) Induced subgraph of the highlighted stopping set with consistent labelling.}
\label{SSvisualH}
\end{figure}

An improved schedule, where both variable nodes and check nodes send messages to each other throughout a single iteration is known by many names including serial scheduling \cite{ZF2002}, \cite{KK2003}, layered scheduling \cite{Hocevar2004} and sequential scheduling \cite{CGW2007}. These algorithms offer an improved decoding performance as information is moving through the Tanner graph more frequently. Examples of decoding algorithms using this scheduling include the max-product algorithm (MPA) \cite{SLG2004} and the belief propagation algorithm (BPA) \cite{CGW2007}. BPA is widely used in LDPC code analysis and is based on the likelihood that a node takes a value given its current value and the values of nearby nodes from previous iterations \cite{Shokrollahi2003}.

Error correction must be implemented differently for each channel. The edge removal algorithm, for example, deals with erasures and thus is not suitable over the BSC. Errors which occur during transmission that are not corrected by the decoding algorithm form what is known as the \textbf{bit error ratio} (BER); the ratio of bits that cannot be corrected versus the total number of bits transmitted \cite{R2003} \cite{Haykin2008}.

In order to test the performance of LDPC codes, we can simulate the transmission of messages over an increasing signal-to-noise ratio (SNR) and calculate the BER of a code under varying conditions. As SNR grows larger, the BER of a code will suddenly decrease depending on the conditions of the channel and the error correcting capability of the LDPC code in use. This curve is known as the \textbf{waterfall} region \cite{R2003}. The best scenario for correcting errors is when the probability of error during transmission over a channel is negligible and when the implemented error correcting code can correct many errors.

\begin{defn} \label{defn:errorfloor} \cite{R2003}
The waterfall region eventually ends in all BER graph curves as anomalous errors cause decoders to fail even with a high SNR ratio. The lowest the BER becomes before levelling is called the \textbf{error floor} of a code.
\end{defn}

The BER is a standard way to analyze the error correcting ability of a code as well as the frame error ratio (FER) \cite{R2003}, \cite{ICV2008}, \cite{Richter2006}. The FER is the ratio of frames or whole messages transmitted which cannot be fully corrected versus the total number of frames transmitted. The largest contributors to the error floor stopping sets and trapping sets.

\section{Cycles and Girth}
\label{CandG}
The decoding method we choose has direct implications for the accuracy and efficiency of decoding. Cycles were the first known negative characteristic of LDPC codes and were extensively studied as they impacted on the accuracy of high performance LDPC codes \cite{TJVW2003}. A \textbf{cycle} in a graph is a sequence of connected nodes which form a closed loop where the initial and final node are the same and no edge is used more than once \cite{BCH2011}. The \textbf{cycle length} is the number of edges a cycle contains, and the length of the smallest cycle in a graph is denoted as its \textbf{girth} \cite{SS2012}.

If no cycles exist within the Tanner graph of a parity-check matrix, then the iterative belief propagation decoding technique is always successful with sufficient iterations \cite{RSU2001}. However, if the neighbours of a node are not conditionally independent then belief propagation methods become inaccurate \cite{TJVW2003}. The inferred solution is to construct a parity-check matrix with no cycles. However, as discussed in Section \ref{chap:Solutions}, this is unessecary as not all cycles negatively impact the decoding efficiency of LDPC codes. In fact, the restriction of girth can lead to constraints on the structure of the code which further impedes the decoding efficiency \cite{TJVW2003}.

The cycles which negatively impact the decoding efficiency of LDPC codes combine to form what are known as stopping sets and trapping sets \cite{DDOV2015}. These sets lead to a high error floor in otherwise efficienct LDPC code constructions throughout various communication channels and affect all high performing decoding algorithms.

\begin{figure*}[!t]
\centering
\subfloat[][]{\includegraphics[width=1.44in]{ER_Tanner}}
\hspace{1pt}\vrule\hspace{1pt}
\subfloat[][]{\includegraphics[width=1.38in]{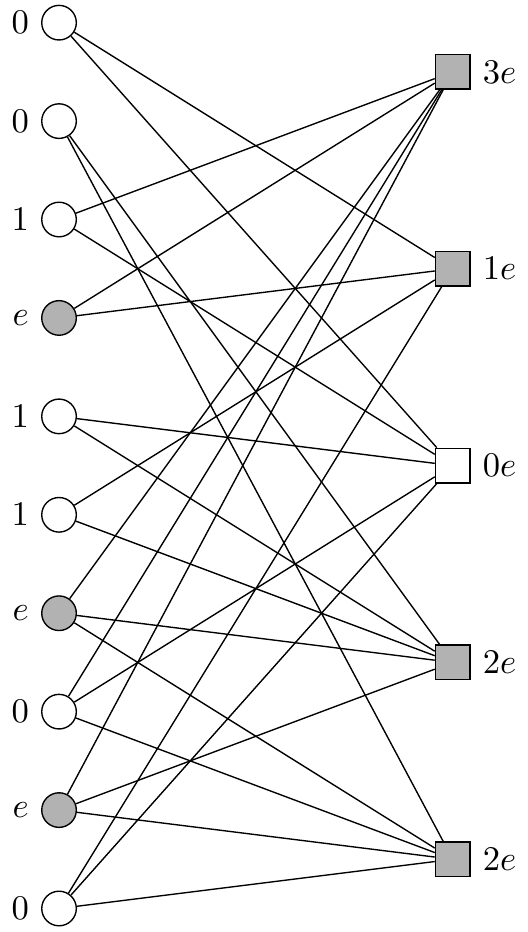}}
\raisebox{18\height}{$\rightarrow$}
\subfloat[][]{\includegraphics[width=1.38in]{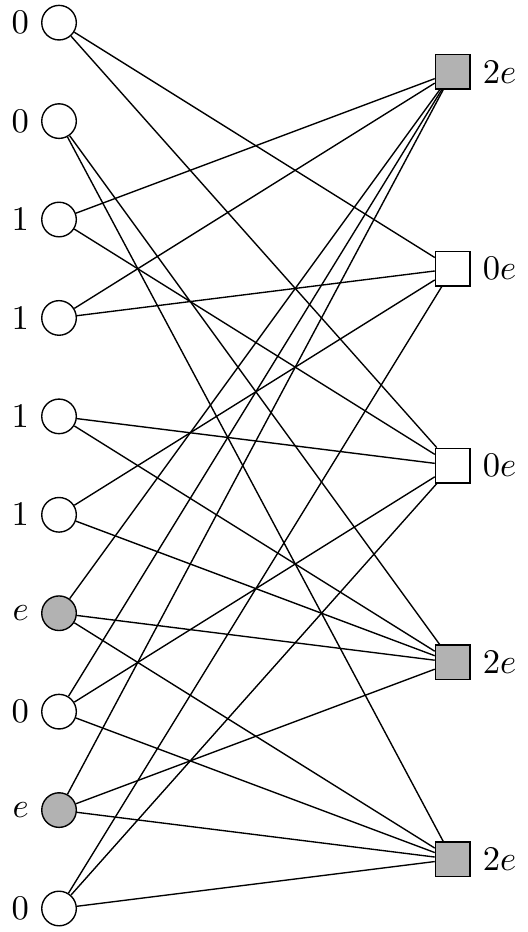}}
\raisebox{18\height}{$\rightarrow$}
\subfloat[][]{\includegraphics[width=1.38in]{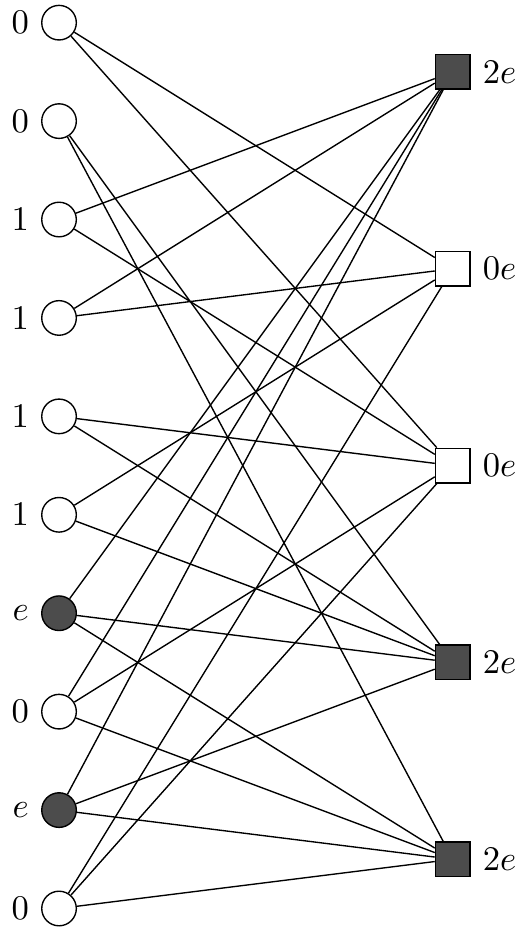}}
\caption{The effect of a stopping set on the ER decoding process for the $5 \times 10$ irregular Tanner graph (a) on the right hand side of the line. (b) shows steps 1 and 2 of the ER algorithm's first iteration. (c) shows the changes made in step 3 and then step 2 of the second iteration. Finally, (d) shows that no further erasures can be corrected, and thus we see that the received vector $v = [0,0,1,e,1,1,e,0,e,0]$ produces a scenario in the Tanner graph where the ER algorithm cannot retrieve the original code word. From Definition \ref{ERalg}, we know that this is due to the presence of a stopping set.}
\label{SSproblem}
\end{figure*}

\section{Stopping Sets over BEC}
\label{SSoverBEC}

Stopping sets are collections of variable and check nodes in the Tanner graph of an LDPC code which greatly reduce its error correcting ability. These sets cause decoding to fail when certain variable nodes are affected by errors after transmission. Stopping sets were first described in 2002 by Di et al \cite{DPTRU2002}, who were researching the average erasure probabilities of bits and blocks over the BEC.

\begin{defn}{\normalfont} \label{defn:SS}\cite{DPTRU2002}
Let $G(H)$ be a Tanner graph and $\mathcal{V}$ be the set of variable nodes in $G(H)$. A \textbf{stopping set}, $\mathcal{S}$, is a subset of $\mathcal{V}$, such that all neighbours of $\mathcal{S}$ are connected to $\mathcal{S}$ at least twice.
\end{defn}

The empty set is also a stopping set and the space of stopping sets is closed under union \cite{DPTRU2002}; if $\mathcal{S}_{1}$ and $\mathcal{S}_{2}$ are both stopping sets then so is $\mathcal{S}_{1} \cup \mathcal{S}_{2}$. The following lemma describes a stopping set by the performance of the LDPC code's decoding algorithm.

\begin{lem} \label{lem: UMSS} \cite{DPTRU2002}
Let $G$ be the generator for an LDPC code over the BEC and $\mathcal{E}$ denote the subset of the set of variable nodes which is erased by the channel after the transmission of a message. Then the set of erasures which remain when the decoder stops is equal to the \textbf{unique maximal stopping set} of $\mathcal{E}$.
\end{lem}

Definition \ref{defn:SS} is now widely accepted \cite{OVZ2005},\cite{RB2013},\cite{RDVW2014}. Given a BEC with erasure probability $\epsilon$, the performance of the code over the BEC is completely determined by the presence of stopping sets \cite{R2003}. Since stopping sets have a combinatorial characterization, their distributions through various Tanner graphs can be analyzed rigorously \cite{DPTRU2002}, \cite{OVZ2005}.

\begin{defn}{\normalfont} \label{def: SN}\cite{OVZ2005}
Let $\mathbb{S}$ denote the collection of all stopping sets in a Tanner graph, $G(H)$. The \textbf{stopping number}, $s^{*}$, of $G(H)$ is the size of the smallest, non-empty stopping set in $\mathbb{S}$.
\end{defn}

The stopping number of a code aids in the analysis of the code's error floor. It is known that the performance of an LDPC code over the BEC is dominated by the small stopping sets in the graph \cite{R2003}. The larger this value is, the lower the error floor of the code. In some cases, this stopping number increases linearly with the number of variable nodes, $|\mathcal{V}|$, in the Tanner graph \cite{OVZ2005}. This can be seen more easily using the stopping ratio.

\begin{defn}{\normalfont} \label{def: SR}\cite{OVZ2005}
Let $G(\mathbf{H})$ be a Tanner graph with $n$ variable nodes and stopping number $s^{*}$. The \textbf{stopping ratio}, $\sigma^{*}$, of a Tanner graph is defined by $s^{*} / n$; the ratio of its stopping number to the number of variable nodes.
\end{defn}

A stopping set in the parity-check matrix of an LDPC code is shown in Example \ref{SSvisual}.

\begin{exmp}
\label{SSvisual}
\cite{Richter2006} Let $\mathcal{C}$ be the code with the following check matrix

\begin{equation*}
\mathbf{H} = 
  \begin{bmatrix}
0 & 0 & 1 & 1 & 0 & 0 & \textbf{1} & 1 & \textbf{1} & 0 \\
1 & 0 & 0 & 1 & 0 & 1 & 0 & 0 & 0 & 1 \\
1 & 0 & 1 & 0 & 1 & 0 & 0 & 1 & 0 & 1 \\
0 & 1 & 0 & 0 & 1 & 1 & \textbf{1} & 0 & \textbf{1} & 0 \\
0 & 1 & 0 & 0 & 0 & 0 & \textbf{1} & 1 & \textbf{1} & 1
  \end{bmatrix}.
\end{equation*}

Columns $7$ and $9$ in $\mathbf{H}$ have been highlighted as they belong to a stopping set. The Tanner graph for $\mathcal{C}$ with the stopping set highlighted is shown in Fig. \ref{SSvisualH}.

A stopping set must be either empty or at least contain two variable nodes. The stopping number, $s^{*}$, of $\mathcal{C}$ is therefore 2 and its stopping ratio, $\sigma^{*}$, $0.2$.
\end{exmp}

An example showing the impact of a stopping set on the decoder is shown in Fig. \ref{SSproblem}, where the edge removal decoding algorithm is used over the BEC.

Solutions to the problem of stopping sets (covered in Section \ref{chap:Solutions}) involve either avoiding or removing small stopping sets in the Tanner graph, leaving only LDPC codes with large stopping sets \cite{RB2013}. While stopping sets are well defined and some solutions exist to minimize their effect of the error floor of LDPC codes, the terminology does not support channels without erasure.

\section{Trapping Sets over BSC, AWGN}
\label{TSoverOther}

Trapping sets, much like stopping sets, are also collections of variable nodes and check nodes which impede the error correcting ability of LDPC code. Only small, elementary trapping sets impact the error floor of LDPC codes over the BSC and AWGNC because of clustering \cite{R2003}, \cite{LM2005}.

The definition of trapping sets came shortly after stopping sets were defined. Similarly to the BEC, when decoding over the BSC and AWGNC, sometimes the maximum iteration count is reached when only a small set of variable nodes are in error. Experiments with the argulis codes lead to the definition of trapping sets \cite{R2003}, \cite{MP2003}.

\begin{defn}{\normalfont} \label{def: FS}\cite{R2003}
Let G($\mathbf{H}$) be a Tanner graph. For a received vector, $y$, of length $n$, we define the \textbf{failure set}, $\mathbf{T}(y)$, to be the set of bits that are not eventually correct using some arbitrary iterative decorder. Decoding is successful on $y$ if and only if $\mathbf{T}(y) = \emptyset$.
\end{defn}

\begin{defn}{\normalfont} \label{def: TS}\cite{LM2005}, \cite{R2003}
If $\mathbf{T}(y) \neq \emptyset$ then $\mathbf{T}(y)$ is a \textbf{trapping set}. More specifically, $\mathbf{T}$ is an \textbf{(a,b) trapping set} in $\mathbf{H}$ if it has $a$ variable nodes for which the sub-graph induced by $\mathbf{T}$ contains $b \geq 0$ odd-degree check nodes.
\end{defn}

Iterative techniques on the BSC and AWGNC distinguish trapping sets from stopping sets over the BEC \cite{R2003}. If there is only iteration by which the decoding algorithm can become trapped then the notion of trapping sets becomes irrelevant.

\begin{figure}[!t]
\centering
\includegraphics[width=2.7in]{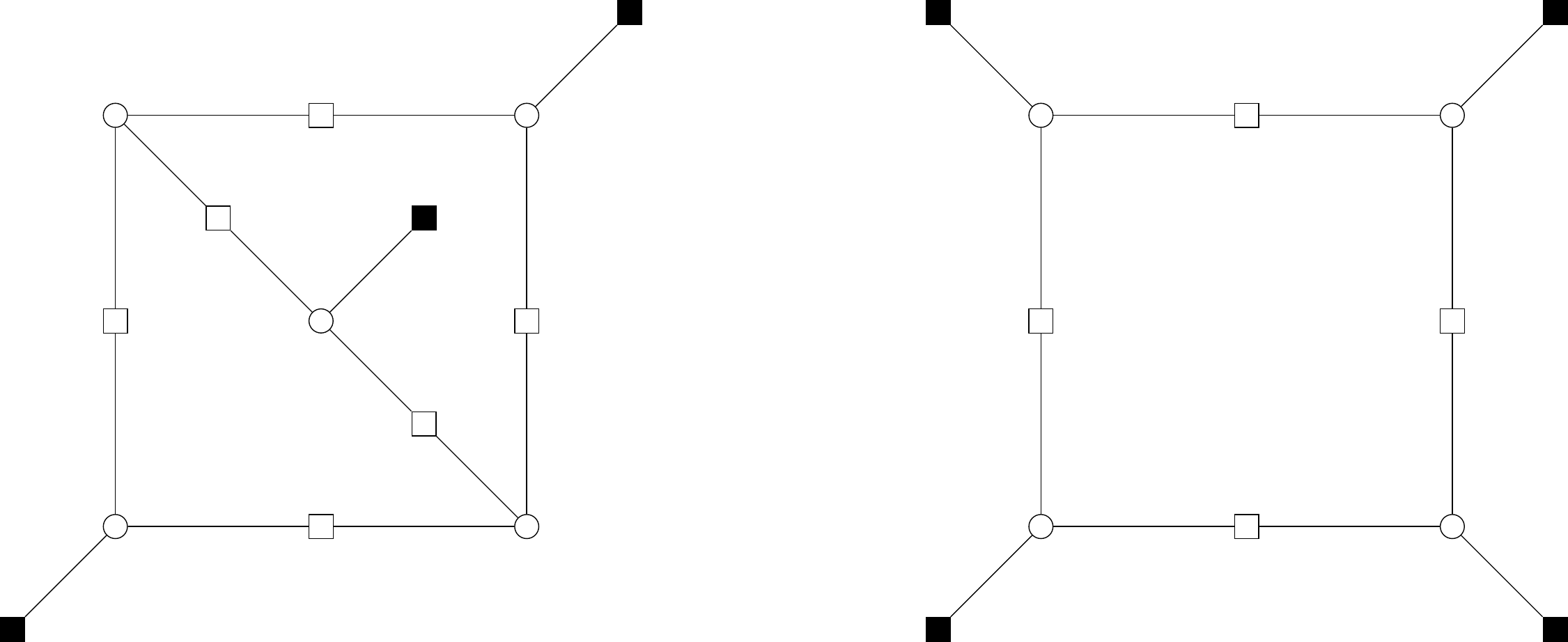}
\caption{A (5,3) trapping set (left) with critical number $k = 3$ and a (4,4) trapping set (right) with critical number $k = 4$ \cite{ICV2008}. These $k$ values are found using the Gallager B decoding algorithm and may vary when other decoding algorithms are applied.}
\label{TSeg}
\end{figure}

\begin{lem} \label{lem: TS0}\cite{R2003}
Let $\mathcal{C}$ be a code using a one-step maximum likelihood decoder, then the trapping sets are precisely the non-zero code words.
\end{lem}

Though, if the channel is BEC, then an iterative decoding failure is said to be due to stopping sets, making stopping sets and trapping sets equivalent over the BEC.

\begin{lem}\label{lem: TS = SS}\cite{R2003}
Let $\mathcal{C}$ be a code using a belief propagation algorithm over BEC, then the trapping sets are precisely the stopping sets.
\end{lem}

Lemma \ref{lem: TS = SS} is an important bridge between trapping sets and stopping sets, allowing us to relate the BEC to the BSC and AWGNC. Decoding failure in an LDPC code over the BSC and AWGNC is largely due to the existence of trapping sets.

Trapping sets pose a real threat to the error correcting ability of LDPC codes; even though there may be very few nodes in error after transmission, if enough of those nodes belong to a trapping set, the decoder will fail.

\begin{defn}{\normalfont}\cite{ICV2008} Let $\mathbf{T}$ be a trapping set. The \textbf{critical number}, $k$, is the minimal number of variable nodes that have to initially be in error for the decoder to become ``trapped" in $\mathbf{T}$.
\end{defn}

It is important to note that the variables nodes that are initially in error do not necessarily belong to the trapping set; it is possible that, at some iteration, the trapping set is entered, causing the decoder to fail. In order to become trapped, the decoder must, after some finite number of iterations, be in error on at least one variable node from $\mathbf{T}$ at every iteration thereafter.

Only trapping sets with a small number of variable nodes and check nodes impact the error-floor of LDPC codes \cite{R2003}, \cite{LM2005}.

\begin{defn}\label{def: STS}{\normalfont}\cite{LM2005}
An $(a,b)$ trapping set in a $[n, k, d]$ code is a \textbf{small trapping set} if $a \leq \sqrt{n}$ and $b \leq 4a$.
\end{defn}

Only these small trapping sets contribute to a larger error-floor \cite{R2003}. Small trapping sets are also of elementary form \cite{LM2005}.

\begin{defn}\label{def: ETS}{\normalfont}\cite{LM2005}
An \textbf{elementary (a,b) trapping set} in a $[n, k, d]$ code is a trapping set for which all check nodes in the induced subgraph have either degree one or two, and there are exactly $b$ degree-one check nodes.
\end{defn}

While check nodes of odd degree larger than one are possible, they are very unlikely within small trapping sets \cite{R2003},\cite{LM2005}. Techniques to find and remove elementary trapping sets have become crucial when constructing high perfoming codes \cite{ICV2008}, \cite{RH2006}.

Two examples of trapping sets \cite{ICV2008} are shown in Fig. \ref{TSeg}. In Fig. \ref{TSeg} the trapping set on the right has a smaller number of variable nodes than the one of the left, however, under the Gallager B decoding algorithm the larger trapping set has a smaller critical number. Thus the performance of the code is limited by the larger trapping set. This idea is quite unintuitive and shows the depth of consideration which must be made when attempting to improve the error floor of LDPC codes.

The problems which trapping sets and stopping sets introduce to LDPC code ares important to research and solve. There do exist methods for constructing LDPC codes by avoiding or removing trapping sets and stopping sets, however, these methods come at the cost of restraining other properties such as code length, density or error correcting ability \cite{JW2004}.

\section{The influence of Stopping Sets and Trapping Sets on LDPC Code Performance}

The original LDPC codes proposed in 1962 by Gallager \cite{G1962}, were construction methods which allowed for varied code rates.

\begin{defn} \cite{MP2003}
A \textbf{Gallager code} is an LDPC code constructed using a parity-check matrix with uniform row weight $i$ and uniform column weight $j$. The code has length $n$ code words and has code rate $R(\mathcal{C})$, which gives a parity-check matrix, $\mathbf{H}$, with $n$ columns and $k$ rows, where $k = n(1-R(\mathcal{C}))$.
\end{defn}

Naive analysis indicated that failed decoding is due to received vectors containing too many errors for the decoding algorithm \cite{G1962}. Analysis of a range of error patterns by Di et al \cite{DPTRU2002} determined that this was not always the case; leading to the definition of stopping sets over the BEC.

A variety of analyses of Gallager codes have shown high performance \cite{MN1996} \cite{MacKay1999}. A construction in 1982 by Margulis \cite{Margulis1982} promised an improved performance over the AWGNC.

For each prime, $p$, let $SL_{2}(p)$ be the Special Linear Group whose elements consist of $2 \times 2$ matrices of determinant $1$ over $\mathbb{Z}_{p}$. This group has $k = (p^{2}-1)(p^{2}-p)/(p-1) = (p^{2}-1)p$ elements. For $p \geq 5$, the Margulis code is of length $n = 2k$, with code rate $R(\mathcal{C}) = 1/2$ \cite{Margulis1982}. The rows of the parity-check matrix are indexed by the elements of $SL_{2}(p)$ and the columns are indexed by two copies of $SL_{2}(p)$; detailed in the following definition.

\begin{defn}\cite{MP2003}\cite{Margulis1982}
Let $SL_{2}(p)$ be generated by the following matrices;

\begin{equation*}
A = 
  \begin{pmatrix}
1 & 2 \\
0 & 1
  \end{pmatrix}
, B = 
  \begin{pmatrix}
1 & 0\\
2 & 1
  \end{pmatrix}
\end{equation*}

If $g \in SL_{2}(p)$ is the index of a row of the parity-check matrix, a one is placed in the columns corresponding to $gA^{2}$, $gABA^{-1}$ and $gB$ on the left hand side of the matrix and also in the columns corresponding to $gA^{-2}$, $gAB^{-1}A^{-1}$ and $gB^{-1}$ on the right hand side of the matrix. This results in a (3,6)-regular parity-check matrix for a \textbf{Margulis code}.
\end{defn}

\begin{figure}[!t]
\centering
\includegraphics[width=3in]{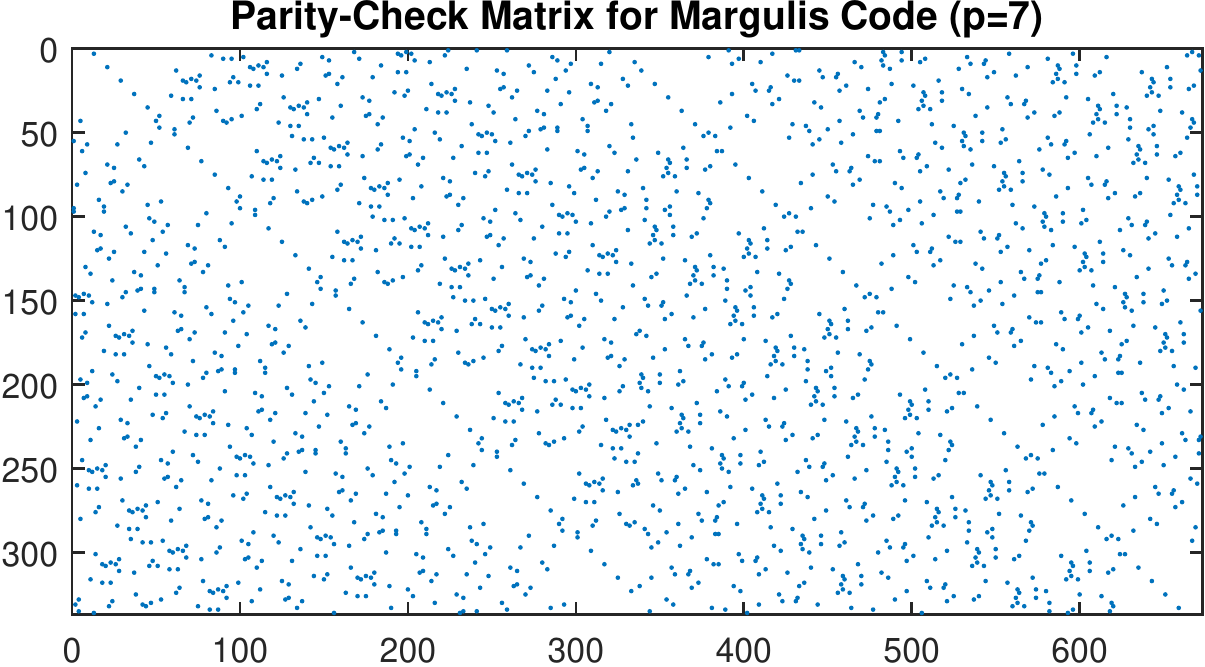}
\caption{Parity-check matrix generated using the Margulis construction, setting $p = 7$ to give a (3,6)-regular $1/2$-rate code with $n=672$. The blue dots represent ones in this matrix with the remaining white space representing zeros.}
\label{MargulisH}
\end{figure}

\begin{figure}[!t]
\centering
\includegraphics[width=3.2in]{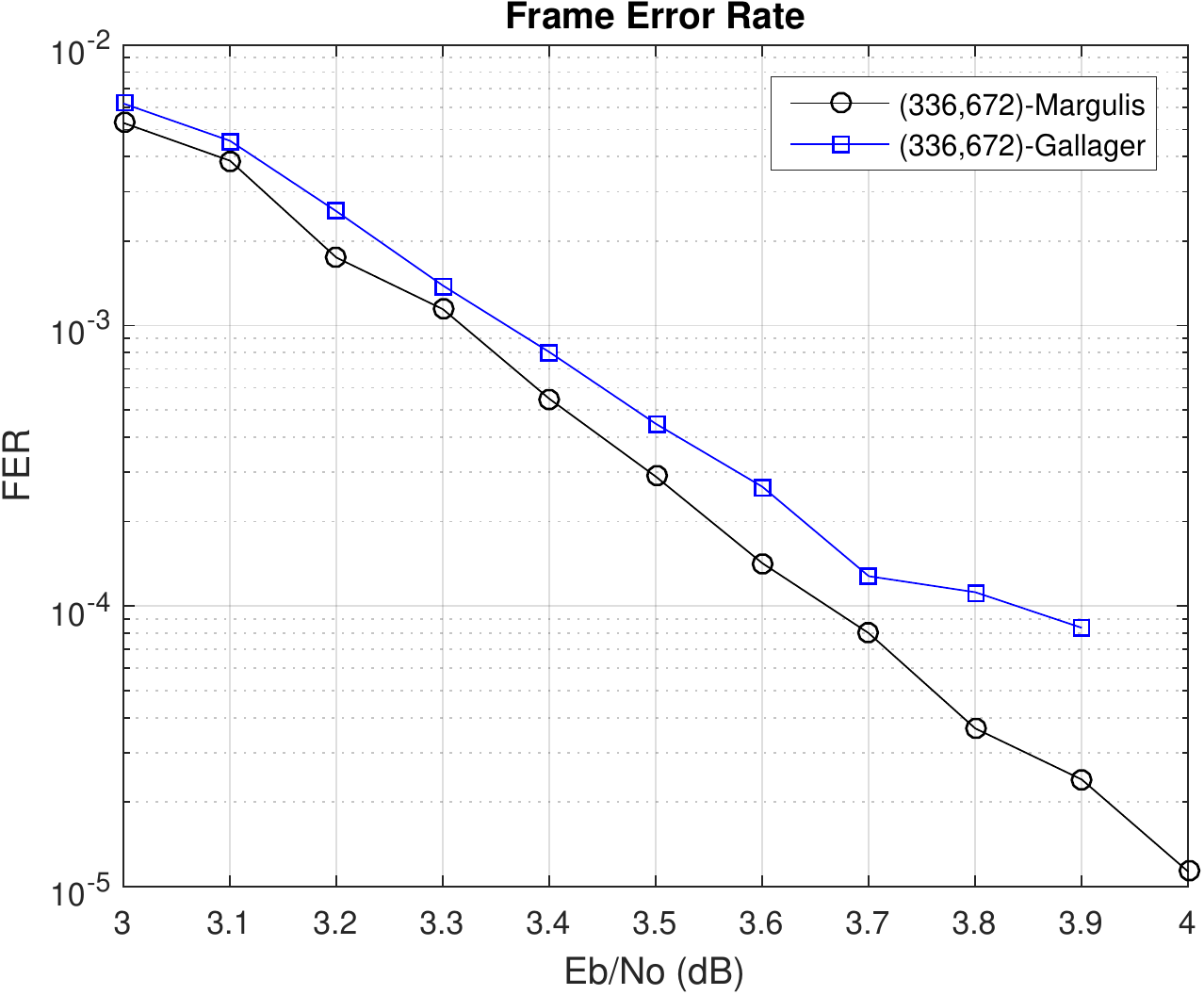}
\caption{BER comparison between the $p=7$ Margulis code (using the same example as presented in Fig. \ref{MargulisH}) and a random Gallager code, both with $n = 672$ and decoded using MPA over the AWGNC. These graphs are also known in the literature as waterfall curves \cite{R2003}.}
\label{MCvGC}
\end{figure}

An example of a parity-check matrix generated using the Margulis construction is shown in Fig. \ref{MargulisH} to demonstrate the sparse nature of LDPC codes. Another example of a Margulis parity-check matrix can be found in \cite{MP2003} where $p = 11$ which corresponds to a $1/2$-rate code with $n=2640$. While the code has a higher performance than a random Gallager code, the error floor is still quite high \cite{MP2003}. This error floor was claimed to be due to near-code words \cite{MP2003}. A comparison between the Margulis code and a random Gallager code, both with $n=672$, can be seen in Fig. \ref{MCvGC}.

\begin{defn} \cite{MP2003}
Let $\mathbf{H}$ be a parity-check matrix. If $\mathbf{x}$ is a vector of weight $w$ and
\begin{equation*}
\mathbf{Hx}^{T}=\mathbf{s}
\end{equation*}
where $\mathbf{s}$ is of weight $v$, then $\mathbf{x}$ is a $(w,v)$ \textbf{near-code word}.
\end{defn}

Near-code words are different from stopping sets. Typical $(w,v)$ near-code words contain $v$ check nodes which are only connected to the variables nodes once. The near-code words in the Margulis code are the $(12,4)$ and $(14,4)$ near-code words \cite{MP2003}.

The high error floors of the Margulis code can be reproduced with a 5 bit approximation to a belief propagation algorithm \cite{R2003}. Near-code words account for 98\% of the error floor performance of the Margulis code. Near code words are trapping sets \cite{R2003}.

Trapping sets are often clustered \cite{R2003}; if one trapping set is found it will often contain nodes which belong to another trapping set. This makes the search for trapping sets somewhat simpler.

Finding both stopping and trapping sets are NP-hard problems \cite{MM2010}, \cite{KS2007}, which makes solutions to these sets difficult to analyse.

The ER decoding algorithm is simple and the effect of stopping sets can be demonstrated easilly. However, decoding over the BSC or AWGNC is much more complex (see Fig. \ref{GA Tanner}). Iterative decoding methods tend to have maximum iteration counts as termination conditions and, as such, demonstrating the effect of trapping sets is difficult to show. In lieu of an example, we remind the reader of the termination conditions for the Gallager A algorithm. This decoder terminates either when all variable nodes send the same values over two consecutive iterations or when some maximum iteration count is reached. In the latter case, the decoder has failed due to the existence of a trapping set.

While we have only discussed the issues with the Margulis code using specific decoding algorithms here, there are many code constructions which contain trapping and stopping sets and decoding algorithms which terminate for the same reasons. For further reading, see \cite{SCRV2006}.

Further reading on stopping sets in LDPC codes include the message passing (MP) algorithm \cite{LMSS2001} and the maximum-likelihood (ML) decoder \cite{PF2004}; both over the BEC. 
Further reading on trapping sets include finite alphabet iterative decoders (FAIDs) \cite{DDPV2011} and constructions based on Latin squares \cite{LM2007}. This construction offers high structure by which stopping sets and trapping sets can be analysed.

If constructions which avoid trapping and stopping sets exist, then the error floors of the associated LDPC codes will lower significantly. This would improve the speed at which almost all digital communication occurs given the already high performance of LDPC codes in modern applications including WiFi \cite{802.11n2009WiFi} and DVB-S2 \cite{302307v1.3.1ETSI2013}.

\section{Current Solutions}
\label{chap:Solutions}
The simple goal is to avoid or completely remove every stopping set or trapping set from an LDPC code. This is both not reasonable given the number of cycles in an LDPC construction \cite{BCH2011} and, more importantly, not necessary \cite{OVZ2005}, \cite{LM2005}.

Only small, elementary trapping sets impact the error floor due to clustering \cite{R2003}, \cite{LM2005}. If there are enough errors in transmission for a decoder to get trapped in a large trapping set then it is highly likely that it would also be trapped in a small trapping set \cite{R2003}. If there are not enough errors for the decoder to become trapped in a large trapping set, the received vector can either be successfully decoded or the decoder will fail due to the presence of at least one small trapping set.

The current solutions to trapping sets are the development of constructions which avoid small trapping sets and the removal of trapping sets from existing constructions.

\begin{figure}[!t]
\centering
\includegraphics[width=3.5in]{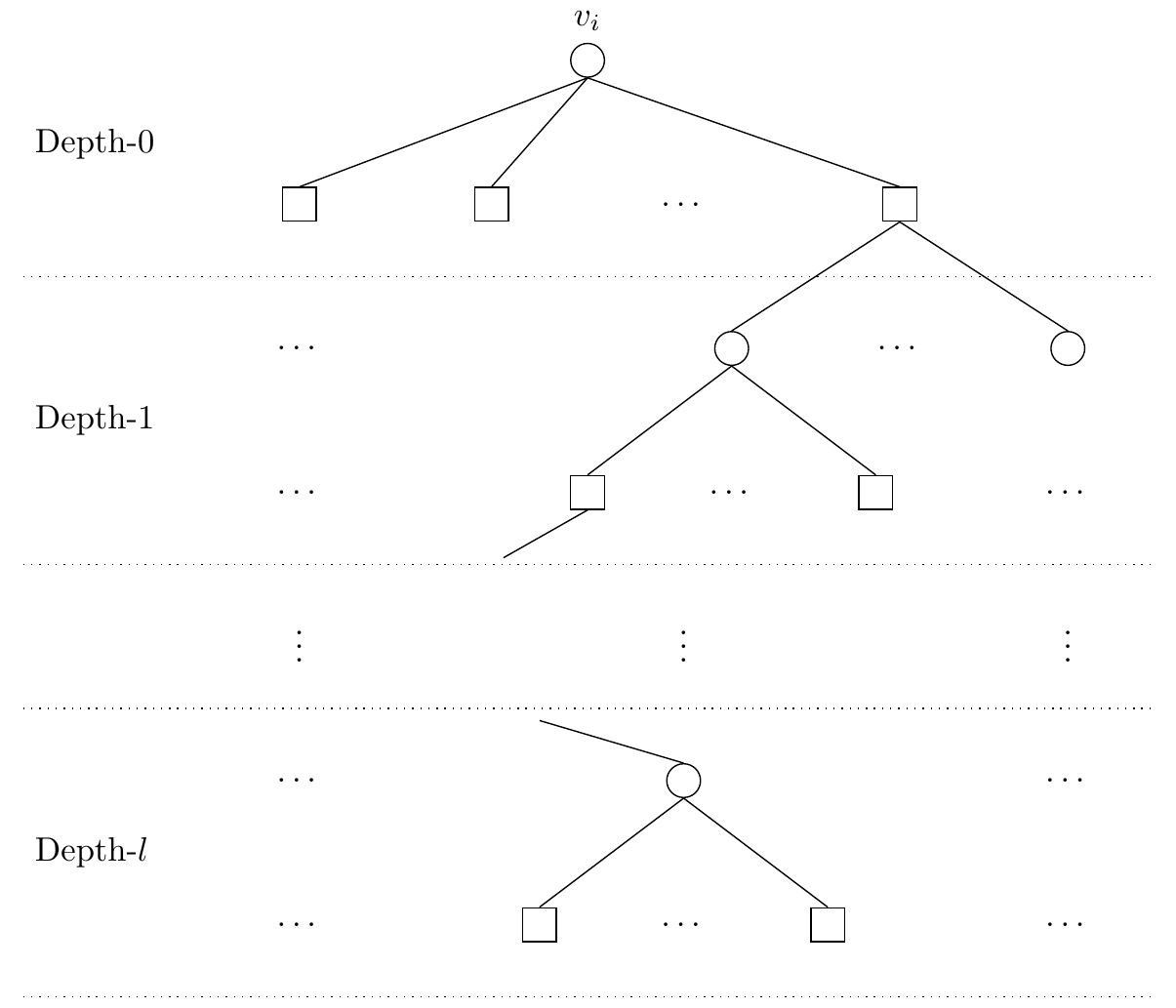}
\caption{\cite{HEA2005} A subgraph (tree) contained in the depth $l$ neighbourhood spreading from the variable node $v_{i}$. Note that $\circ$ here represents a variable node and $\square$ represents a check node.}
\label{TreeBF}
\end{figure}

\subsection{Avoiding Trapping Sets}
Stopping sets pose threats to the error correction of messages sent over the BEC, however, in practice the AWGNC is used and so, when discussing proposed solutions, we focus on the influence of trapping sets over the AWGNC.

The progressive-edge-growth (PEG) construction \cite{HEA2001}, \cite{HEA2005} is a method of constructing Tanner graphs with high girth. Many trapping sets include small cycles, so the likelihood of a small trapping set being constructed is small with a graph of high girth \cite{MSW2007}. In order to give the definition for the PEG construction, some definitions are needed. The PEG construction method uses variable and check node degree sequences \cite{HEA2005}. The \textbf{variable node degree sequence} is denoted
\begin{align*}
D_{v} &= \{d_{v_{0}}, d_{v_{1}}, \dots, d_{v_{n-1}}\},
\intertext{where $d_{v_{i}}$ is the degree of variable node $v_{i}$, $ 0 \leq i \leq n-1$, and $d_{v_{0}} \leq d_{v_{1}} \leq \dots \leq d_{v_{n-1}}$. The \textbf{parity check sequence} is denoted}
D_{c} &= \{d_{c_{0}}, d_{c_{1}}, \dots, d_{c_{m-1}}\},
\end{align*}
where $d_{c_{j}}$ is the degree of check node $c_{j}$, $ 0 \leq j \leq m-1$, and $d_{c_{0}} \leq d_{c_{1}} \leq \dots \leq d_{c_{m-1}}$.

The construction partitions the set of edges, $E$, to
\begin{align*}
E = E_{v_{0}} \cup E_{v_{1}} \cup \dots \cup E_{v_{n-1}},
\end{align*}
where $E_{v_{i}}$ contains all edges incident on symbol node $v_{i}$. The $k^{th}$ edge incident on $v_{i}$ is denoted as $E_{v_{i}}^{k}$, where $0 \leq k \leq d_{v_{i}}-1$.

The \textbf{neighbourhood} of depth $l$ for variable node $v_{i}$ is $\mathcal{N}_{v_{i}}^{l}$ and is defined as the set of all check nodes included in a subgraph (tree) spreading from variable node $v_{i}$ within depth $l$. This is demonstrated in Fig. \ref{TreeBF}. The complement of $\mathcal{N}_{v_{i}}^{l}$ is $\mathcal{N}_{v_{i}}^{-l} = C \backslash \mathcal{N}_{v_{i}}^{l}$, where C is the set of check nodes. The subgraph (tree) generated this way is constructed breadth-first with $v_{i}$ as the root. Given the parameters $n$, $m$ and $D_{v}$, we define the PEG construction as follows.

\textbf{Progressive edge-growth algorithm (PEG)} \cite{HEA2005}: \\
\textbf{for} $i = 0$ to $n - 1$ do\\
\textbf{begin}\\
\tab \textbf{for} $k = 0$ to $d_{v_{i}} - 1$ do\\
\tab \textbf{begin}\\
\tab \tab \textbf{if} $k = 0$\\
\tab \tab \tab $E_{v_{i}}^{0} \leftarrow$ edge ($c_{j}$, $v_{i}$), where $e_{v_{i}}^{0}$ is the first edge\\
\tab \tab \tab incident to $v_{i}$ and $c_{j}$ is a check node that it\\
\tab \tab \tab has the lowest check-node degree in\\
\tab \tab \tab $E_{v_{0}} \cup E_{v_{1}} \cup \dots \cup E_{v_{i-1}}$.\\
\tab \tab \textbf{else}\\
\tab \tab \tab Expand a subgraph from symbol node $v_{i}$ up to\\
\tab \tab \tab depth $l$ in $E_{v_{0}} \cup E_{v_{1}} \cup \dots \cup E_{v_{i-1}}$ until the\\
\tab \tab \tab cardinality of $\mathcal{N}_{v_{i}}^{l}$ stops increasing but is less\\
\tab \tab \tab than $m$, or $\mathcal{N}_{v_{i}}^{-l} \neq \varnothing$ but $\mathcal{N}_{v_{i}}^{l+1} = \varnothing$, then $E^{k}_{v_{i}} \leftarrow$\\
\tab \tab \tab edge ($c_{j}$, $v_{i}$), where $E^{k}_{v_{i}}$ is the $k^{th}$ edge incident\\
\tab \tab \tab to $v_{i}$ and $c_{j}$ is a check node chosen from the\\
\tab \tab \tab set $\mathcal{N}_{v_{i}}^{-l}$ having the lowest check-node degree.\\
\tab \tab \textbf{end}\\
\tab \textbf{end}\\
\textbf{end}

\begin{figure*}[!t]
\centering
\subfloat[][]{\includegraphics[width=1.44in]{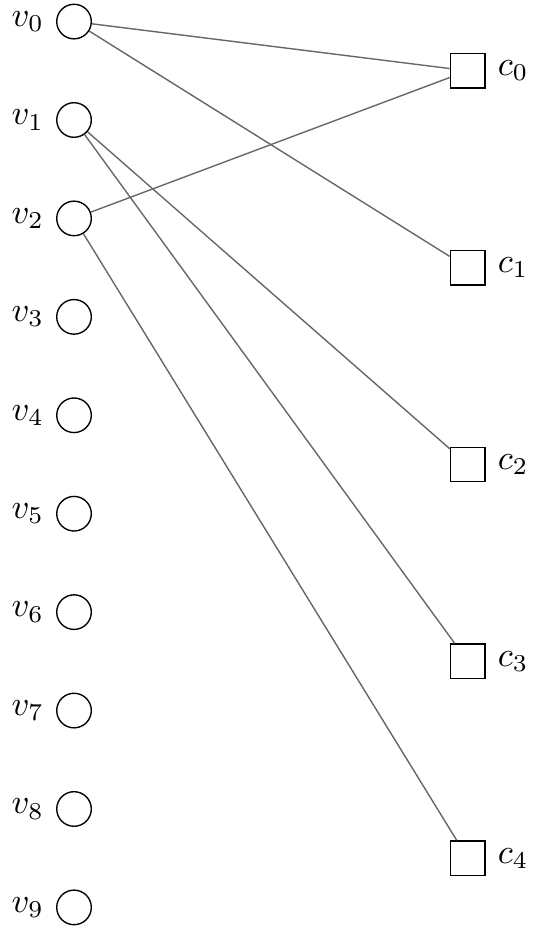}}
\raisebox{18\height}{$\rightarrow$}
\subfloat[][]{\includegraphics[width=1.44in]{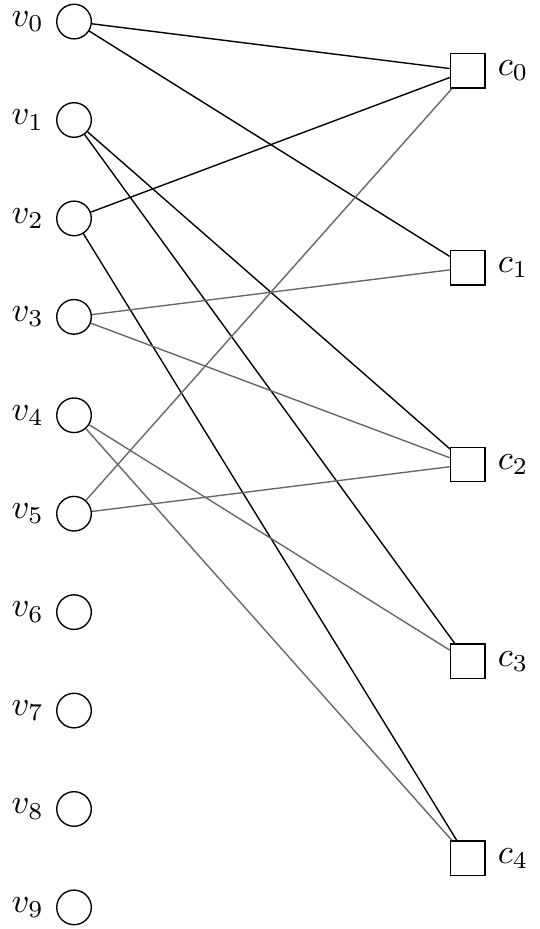}}
\raisebox{18\height}{$\rightarrow$}
\subfloat[][]{\includegraphics[width=1.44in]{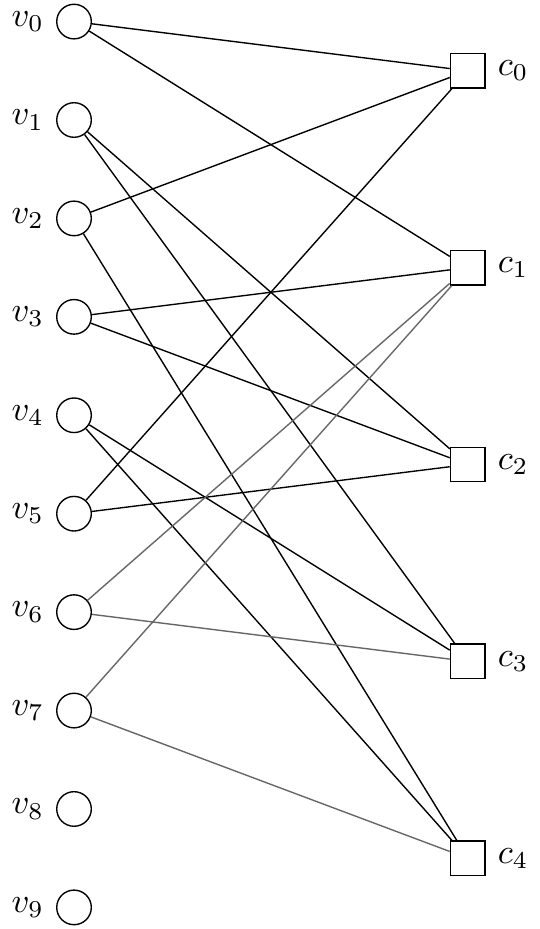}}
\raisebox{18\height}{$\rightarrow$}
\subfloat[][]{\includegraphics[width=1.44in]{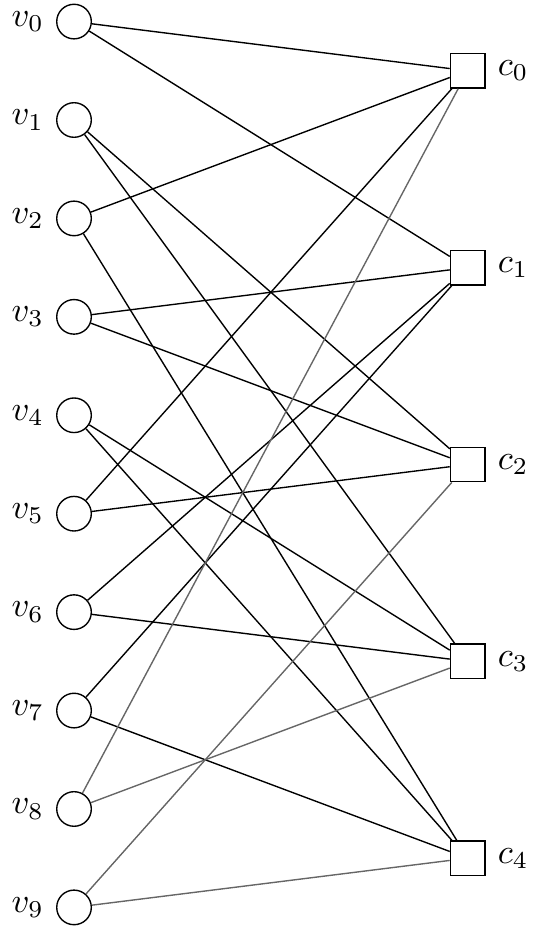}}
\caption{A PEG construction with $n = 10$, $m = 5$ and $D_{v} = \{2,2,2,2,2,2,2,2,2,2\}$. Check nodes are chosen based on index order. The first edge chosen for each variable node is chosen from the check nodes with lowest degree at random. The generation of the subgraphs and subsequent edge placement are the factors which highlight this construction method. In (a), the edge choices are simplistic as the breadth-first subgraphs are of low depth. This decision making continues in (b) until $s_{5}$ is considered, where the edge choice is restricted to $c_{2}$ or $c_{3}$ due to connections in the subgraph to check nodes $c_{1}$ and $c_{4}$. These choices can be observed in both remaining figures (c) and (d). One notable choice is the $2^{nd}$ edge decision for $v_{9}$ in (d), where the only remaining option once $c_{2}$ is chosen becomes $c_{4}$, which gives a uniform check node degree sequence.}
\label{PEGex}
\end{figure*}

When presented with check nodes of the same degree, a decision must be made; the selection of a check node at random or the selection of a check node according to some order. An improved construction, the randomized-PEG construction \cite{VDP2008}, chooses at random, though the deterministic nature of the ordered check node process might be of use \cite{HEA2005}. An example of PEG construction is given in Fig. \ref{PEGex}, setting $n = 10$, $m = 5$ and $D_{v} = \{2,2,2,2,2,2,2,2,2,2\}$.

The PEG construction maximises the local girth of a variable node when a new edge is added to the node \cite{HEA2005}. After the discovery of stopping and trapping sets the PEG construction was modified \cite{DDOV2015}, \cite{VDP2008}.

The PEG construction is notable for its ability to create high girth LDPC codes, however, the number of cycles is not controlled. Trapping sets are formed by a combination of several cycles \cite{DDOV2015}. The PEG algorithm, while having a higher girth than aternate constructions, contains more trapping sets; thus leaving the error floor open to improvement.

The RandPEG construction improves upon the PEG algorithm by minimizing cycles at the same time as reducing the computational complexity of the PEG algorithm \cite{VDP2008}. This can be further improved by adding an objective function to avoid small trapping sets \cite{DDOV2015}. Quasi-cyclic LDPC (QC-LDPC) codes, which are used in many applications \cite{2DDOV2016}, \cite{WYD2008} can be constructed using the Improved RandPEG algorithm \cite{DDOV2015}. The objective function used in the improved RandPEG algorithm detects all (5,3) and (6,4) trapping sets, removing all (5,3) trapping sets and as many (6,4) trapping sets as possible without adversely affecting the performance of the LDPC code. The characterization of trapping sets is achieved in \cite{DDOV2015} through the locations of check nodes in different levels of depth-$l$ trees (see Fig. \ref{TreeBF}). The resulting construction is as follows:

\textbf{Improved RandPEG algorithm (RPEG)} \cite{DDOV2015}: \\
\textbf{for} $i = 0$ to $n - 1$ do\\
\textbf{begin}\\
\tab \textbf{for} $k = 0$ to $d_{v_{i}} - 1$ do\\
\tab \textbf{begin}\\
\tab \tab \textbf{if} $k = 0$\\
\tab \tab \tab $E_{v_{i}}^{0} \leftarrow$ edge ($c_{j}$, $v_{i}$), where $e_{v_{i}}^{0}$ is the first edge\\
\tab \tab \tab incident to $v_{i}$ and $c_{j}$ is a check node such that\\
\tab \tab \tab it has the lowest check-node degree in\\
\tab \tab \tab $E_{v_{0}} \cup E_{v_{1}} \cup \dots \cup E_{v_{i-1}}$.\\
\tab \tab \textbf{else}\\
\tab \tab \tab Expand a subgraph from symbol node $v_{i}$ up to\\
\tab \tab \tab depth $l$ in $E_{v_{0}} \cup E_{v_{1}} \cup \dots \cup E_{v_{i-1}}$ until the\\
\tab \tab \tab cardinality of $\mathcal{N}_{v_{i}}^{l}$ stops increasing but is less\\
\tab \tab \tab than $m$, or $\mathcal{N}_{v_{i}}^{-l} \neq \varnothing$ but $\mathcal{N}_{v_{i}}^{l+1} = \varnothing$. Remove from\\
\tab \tab \tab $\mathcal{N}_{v_{i}}^{-l}$ all check nodes that appear at least once in\\
\tab \tab \tab the depth-$3$ tree spreading from $v_{i}$. This\\
\tab \tab \tab removes all check nodes that would create cycles\\
\tab \tab \tab  of size $< 8$.\\
\tab \tab \tab \textbf{for} $c_{m}$ in $\mathcal{N}_{v_{i}}^{-l}$ do\\
\tab \tab \tab \tab Compute the number of (5,3) and (6,4) trap-\\
\tab \tab \tab \tab ping sets that would be created if $c_{m}$ is\\
\tab \tab \tab \tab selected. Remove all check nodes that would\\
\tab \tab \tab \tab create (5,3) trapping sets and remove check\\
\tab \tab \tab \tab nodes which create more than the smallest\\
\tab \tab \tab \tab number of (6,4) trapping sets.\\
\tab \tab \tab \tab \textbf{If} $\mathcal{N}_{v_{i}}^{-l} \neq \varnothing$\\
\tab \tab \tab \tab \tab $E^{k}_{v_{i}} \leftarrow$ edge $(c_{m},v_{i})$, where $E^{k}_{v_{i}}$ is the $k^{th}$\
\tab \tab \tab \tab \tab edge incident to $v_{i}$ and $c_{m}$ is a check node\\
\tab \tab \tab \tab \tab chosen from the remaining nodes in $\mathcal{N}_{v_{i}}^{-l}$.\\
\tab \tab \tab \tab \textbf{else}\\
\tab \tab \tab \tab \tab Declare a design failure.\\
\tab \tab \tab \tab \textbf{end}\\
\tab \tab \tab \textbf{end}\\
\tab \tab \textbf{end}\\
\tab \textbf{end}\\
\textbf{end}

The Improved RandPEG construction algorithm, while having a high computational complexity, performs the task of avoiding trapping sets optimally for given dimensions of an LDPC code. Possible improvements to this construction method include lowering the computational complexity and potentially lowering the girth. The removal of all cycles is unnecessary as not all cycles contribute to trapping sets \cite{OVZ2005}, \cite{LM2005}. The inclusion of a lower girth into a construction which also contains no small trapping sets could lead to an LDPC code with a higher decoding performance.

\subsection{Removing Stopping and Trapping Sets}
The performance of LDPC codes is constrained by the presence of cycles and trapping sets within the code's parity-check matrix. We discuss two methods of removing trapping sets; the addition of a redundant parity-check equation \cite{LHMH2006} and the use of Tanner graph covers \cite{ICV2008}.

\subsection*{Redundant Parity-Check Equations}
Adding a redundant parity-check equation is equivalent to adding a redundant row to the parity-check matrix. This has been used in an attempt to remove the trapping sets present in the [2640, 1320] Margulis code \cite{LHMH2006}. The $(12,4)$ and $(14,4)$ trapping sets in the $[2640, 1320]$ Margulis code are elementary point trapping sets \cite{MP2003}. \textbf{Point trapping sets} are subsets of variable nodes that contain all errors ever to occur throughout the decoding process \cite{LHMH2006}.

A redundant parity check row is identified which, when added to the parity-check matrix, potentially disrupts the (12,4) and (14,4) elementary trapping sets. This parity-check row is identified through a genie-aided random search which relies on information about trapped variables not available during decoding \cite{LHMH2006}. As random searches cannot be used in applied error correction a structured search was considered to be more useful. The structured search identifies variable and check nodes which connect to both the (12,4) and (14,4) trapping sets and combines the projection of the involved nodes such that a redundant parity-check row can be added to eliminate the effect of those trapping sets.

\begin{figure}[!t]
\centering
\includegraphics[width=3.3in]{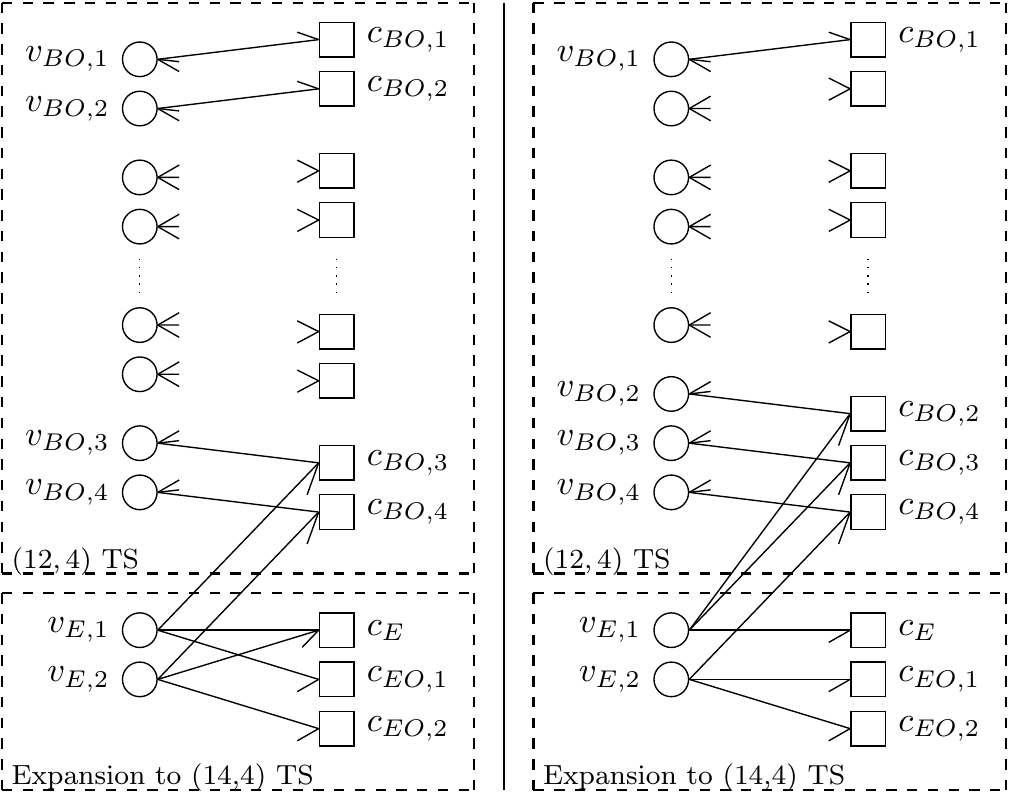}
\hspace*{-190pt}\mbox{\subfloat[][]{}}\hspace{190pt}\hspace*{-65pt}\mbox{\subfloat[][]{}}\hspace{65pt}
\caption{\cite{LHMH2006} The trapping set structure of a $(12,4)$ trapping set and its $(14,4)$ expansion configurations; (a) left and (b) right. For (a), the two expansion variables are denoted $v_{E,1}$ and $v_{E,2}$ and the check node connected to both of the variables nodes is denoted $c_{E}$. The unsatisfied check nodes in the original $(12,4)$ trapping set are denoted $c_{BO,1}$ through $c_{BO,4}$ and the variable nodes connected to these check nodes are denoted $v_{BO,1}$ through $v_{BO,4}$. The check nodes of degree one in the expansion of the trapping set are denoted $c_{EO,1}$ and $c_{EO,2}$. The node labels for (b) follow similarly.}
\label{TSext}
\end{figure}

The only way to disrupt the (12,4) and (14,4) trapping sets in the Margulis code is if the projection of both the 12 variables and the 14 variables on the redundant parity-check equation has row weight one \cite{VDP2008}. This can be most reliably achieved by extending (12,4) trapping sets to (14,4) trapping sets (see Fig. \ref{TSext}). Given that the Margulis code has a (3,6)-regular parity-check matrix, an elementary (a,b) trapping set contains a fixed number of check nodes. Let $e$ denote the number of check nodes connected to two variable nodes within the trapping set, then $e = (3a-b)/2$ and therefore two variable nodes and three check nodes must be added to extend a (12,4) trapping set to a (14,4) trapping set \cite{VDP2008}. At most one check node can be connected to both of the added variable nodes such that 4-cycles are not created. Such an extension which avoids the creation of 4-cycles in the Margulis code is only possible in two configurations.
\begin{enumerate}
\item[(a)] The two degree-one check nodes of the basic (12,4) trapping set are connected through two additional variable nodes to one additional check node.
\item[(b)] In the second configuration, the additional variable nodes do not share a check node.
\end{enumerate}
These configurations are demonstrated in Fig. \ref{TSext}. The existing check and variable nodes neighbouring the additional check and variable nodes are linearly combined to generate a redundant parity-check equation. A structured search is then used to ensure that this projection has row weight one in both the (12,4) and (14,4) trapping set.

The addition of a redundant parity-check equation focuses on point elementary trapping sets from the Margulis code, where the structure of the trapping sets are well known. If this method were applied to other LDPC codes, the location and structure of the trapping sets within such a code are unknown. The redundant rows are computationally inexpensive to compute, however, the code rate of the resulting LDPC code will be reduced and the extra row increases the number of operations per decoding iteration, though by a negligible amount. Another potential problem is the success rate of this solution. The addition of a redundant parity-check equation does not guarantee that the trapping set will be disrupted \cite{LHMH2006}.

\subsection*{Tanner Graph Covers}
Another method capable of eliminating trapping sets is the utilization of graph covers \cite{ICV2008}. This method constructs an LDPC code $\mathcal{C}^{(2)}$ of length $2n$ given a code $\mathcal{C}$ of length $n$. The parity check matrix of this code is denoted $H^{(2)}$ and is initialized to

\begin{equation*}
H^{(2)} = 
\begin{bmatrix}
    H       & 0 \\
    0       & H
\end{bmatrix}
\end{equation*}

The operation of changing the value of $H^{(2)}_{t,k}$ and $H^{(2)}_{m+t,n+k}$ to ``$0$", and $H^{(2)}_{m+t,k}$ and $H^{(2)}_{m,n+k}$ to ``$1$" is termed as edge swapping $e$. The graph covers method requires that the locations of dominant trapping sets are known. The method of edge swapping is then described as follows.

\textbf{Graph covers algorithm} \cite{ICV2008}:
\begin{enumerate}
\item Take two copies, $\mathcal{C}_{1}$ and $\mathcal{C}_{2}$, of the code $\mathcal{C}$. Since the codes are identical they share the same trapping sets. Initialize $SwappedEdges=\varnothing$, $FrozenEdges=\varnothing$;
\item Order the trapping sets by their critical numbers.
\item Choose a trapping set $\mathbf{T}_{1}$ in the Tanner graph of $\mathcal{C}_{1}$, with minimal critical number. Let $E_{\mathbf{T}_{1}}$ denote the set of all edges in $\mathbf{T}_{1}$. If $E_{\mathbf{T}_{1}} \cap SwappedEdges \neq \varnothing$ go to step 5, else go to step 4.
\item Swap an arbitrarily chosen edge $e \in E_{\mathbf{T}_{1}} \textbackslash$ $FrozenEdges$. Set $SwappedEdges = SwappedEdges \cup e$.
\item ``Freeze" the edges $E_{\mathbf{T}_{1}}$ from $\mathbf{T}_{1}$ so that they cannot be swapped in the following steps. Set $FrozenEdges = FrozenEdges \cup E_{\mathbf{T}_{1}}$.
\item Repeat steps 2 to 4 until all trapping sets of the desired size are removed.
\end{enumerate}

Possible improvements to the graph cover method are to prioritize specific edges for swapping and freezing to avoid creating trapping sets of the same critical number \cite{ICV2008}. However, experimentally, all trapping sets with minimal critical number were removed using the above algorithm \cite{ICV2008}. The graph covers method gave improved FER results for a Tanner code \cite{TSF2001}, a Margulis code \cite{RV2000} and a MacKay code \cite{MacKay2005} using the Gallager B decoding algorithm. While the decoding method was constant throughout these results, the application of graph covers will optimize FER performance using an arbitrary decoding algorithm \cite{ICV2008}.

The LDPC code $\mathcal{C}^{(2)}$ created from code $\mathcal{C}$ has code rate $r^{(2)} \leq r$ and minimum distance $2d \geq d^{(2)} \geq d$. An increase to the minimum distance of the code $\mathcal{C}^{(2)}$ gives higher error correcting capabilities than $\mathcal{C}$. However, a lower code rate could decrease the overall efficiency. The lower row and column weight of $H^{(2)}$ gives $\mathcal{C}^{(2)}$ higher FER performance than $\mathcal{C}$, though with a trade-off of low decoding complexity.

The trade-offs associated with removing trapping sets are more severe than the surveyed construction methods which avoid them (code length, decoding speed from check nodes, etc). The current research goal remains the creation of a construction or modifiable construction which can either avoid or remove small elementary trapping sets without penalty to the code's error correcting ability or decoding efficiency.

\section{Conclusion}
Throughout this survey we have covered the literature surrounding LDPC codes, communication channels and decoding techniques. The negative impact cycles have on LDPC code efficiency is noted and the problem of stopping sets and trapping sets have been defined and discussed including the dominance of small elementary trapping sets over AWGNC. A small variety of partial solutions such as the randomized progressive edge-growth algorithm and Tanner graph covers are discussed. The research goal remains to find constructions of LDPC codes without small trapping sets.


%

\appendices


\section*{Acknowledgment}

The authors would like to thank Professor Ian Turner, who worked closely with us throughout our research, Emeritus Professor Ed Dawson and Dr Harry Bartlett for their help in the final stages before submission, Dr Dhammika Jayalath for his help with the decoding simulations over AWGNC, and Xuan He for his suggestions on how to present our BER data. Computational resources and services used in this work were provided by the HPC and Research Support Group, Queensland University of Technology, Brisbane, Australia. A. Price is supported by an APA Scholarship.

\ifCLASSOPTIONcaptionsoff
  \newpage
\fi



%

%

\begin{IEEEbiography}[{\includegraphics[width=1in,height=1.25in,clip,keepaspectratio]{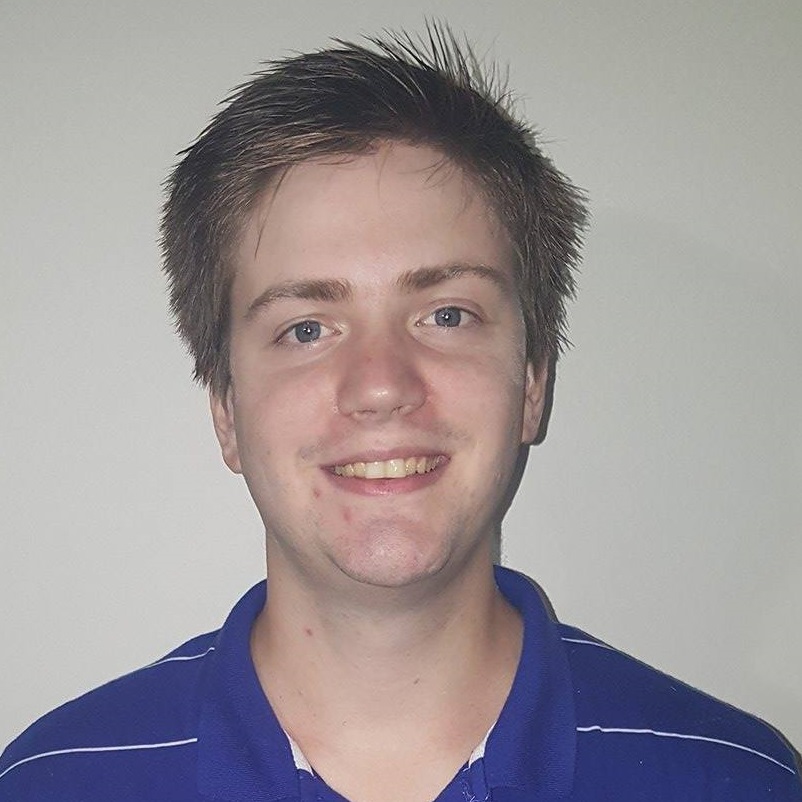}}]{Aiden Price}
was awarded the QUT Dean's scholarship and received a BSc and $1^{st}$ class honours in mathematics from Queensland University of Technology. Aiden is currently studying a PhD at QUT in the school of mathematics under the supervision of Doctor Harry Bartlett and Emeritus Professor Ed Dawson and is funded by the APA scholarship. His research interests are coding theory and its application to digital communications and cryptography.
\end{IEEEbiography}

\begin{IEEEbiography}[{\includegraphics[width=1in,height=1.25in,clip,keepaspectratio]{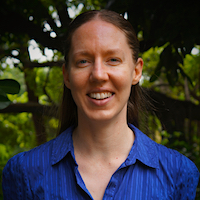}}]{Joanne Hall}
 received a BSc and MPhil in mathematics from the Australian National University. She graduated with a PhD from RMIT University
in 2011 under the supervision of Asha Rao in
the Information Security and Informatics research
group. 
Dr Hall spent one year as a postdoctoral research scientist at Charles University in Prague and four years as a Lecturer at the Queensland University of Technology in Brisbane.  In 2017 she has returned to RMIT University as a Lecturer in the School of Science.  Her research interests are algebraic and combinatorial  structures and their applications in digital communication.
\end{IEEEbiography}





\end{document}